\documentclass[iop,tighten]{emulateapj}

\begin{document}

\shorttitle{Resolved Transition Disk Cavities}

\shortauthors{Andrews et al.}

\title{Resolved Images of Large Cavities in Protoplanetary Transition Disks}

\author{Sean M. Andrews\altaffilmark{1}, David J. Wilner\altaffilmark{1}, Catherine Espaillat\altaffilmark{1,2}, A. M. Hughes\altaffilmark{3}, \\ C. P. Dullemond\altaffilmark{4}, M. K. McClure\altaffilmark{5}, Chunhua Qi\altaffilmark{1}, and J. M. Brown\altaffilmark{1}}
\altaffiltext{1}{Harvard-Smithsonian Center for Astrophysics, 60 Garden Street, Cambridge, MA 02138}
\altaffiltext{2}{NSF Astronomy \& Astrophysics Postdoctoral Fellow}
\altaffiltext{3}{Department of Astronomy, University of California at Berkeley, 601 Campbell Hall, Berkeley, CA 94720}
\altaffiltext{4}{Institut f{\"u}r Theoretische Astrophysik, Universit{\"a}t Heidelberg, Albert-Ueberle-Str.~2, Heidelberg, Germany 69120}
\altaffiltext{5}{Department of Astronomy, University of Michigan, 830 Dennison Bldg, 500 Church Street, Ann Arbor, MI 48109}

\begin{abstract}
Circumstellar disks are thought to experience a rapid ``transition" phase in 
their evolution that can have a considerable impact on the formation and early 
development of planetary systems.  We present new and archival high angular 
resolution ($0\farcs3 \approx 40$-75\,AU) Submillimeter Array (SMA) 
observations of the 880\,$\mu$m (340\,GHz) dust continuum emission from 12 such 
transition disks in nearby star-forming regions.  In each case, we directly 
resolve a dust-depleted disk cavity around the central star.  Using 
two-dimensional Monte Carlo radiative transfer calculations, we interpret these 
dust disk structures in a homogeneous, parametric model framework by 
reproducing their SMA continuum visibilities and spectral energy 
distributions.  The cavities in these disks are large ($R_{\rm cav} = 
15$-73\,AU) and substantially depleted of small ($\sim$$\mu$m-sized) dust 
grains, although their mass contents are still uncertain.  The structures of 
the remnant material at larger radii are comparable to normal disks.  We 
demonstrate that these large cavities are relatively common among the 
millimeter-bright disk population, comprising at least 1 in 5 (20\%) of the 
disks in the bright half (and $\ge$26\%\ of the upper quartile) of the 
millimeter luminosity (disk mass) distribution.  Utilizing these results, we 
assess some of the physical mechanisms proposed to account for transition disk 
structures.  As has been shown before, photoevaporation models do not produce 
the large cavity sizes, accretion rates, and disk masses representative of this 
sample.  It would be difficult to achieve a sufficient decrease of the dust 
optical depths in these cavities by particle growth alone: substantial growth 
(to meter sizes or beyond) must occur in large (tens of AU) regions of low 
turbulence without also producing an abundance of small particles.  Given those 
challenges, we suggest instead that the observations are most commensurate with 
dynamical clearing due to tidal interactions with low-mass companions -- 
young brown dwarfs or giant planets on long-period orbits. 
\end{abstract}

\section{Introduction}

Planets are made from gas and dust in disks that orbit young stars.  The 
physical characteristics and evolution of that disk material have a profound 
impact on the outcome and efficiency of the planet formation process.  While a 
massive theoretical effort has been made to elucidate the physical mechanisms 
that contribute to the assembly of a planetary system, any successful planet 
formation model needs to be informed by observations.  An important goal is to 
generate models that incorporate empirical insights on the properties of 
circumstellar disks and are able to reproduce key demographic features of the 
exoplanet population \citep[e.g.,][]{ida04,ida05,alibert05,mordasini09}.  One 
essential aspect of meeting that goal is the ability to make measurements 
related to the initial conditions imposed by the disk material.  However, those 
constraints on initial conditions must be coupled with a characterization of 
disk evolution, which is ultimately responsible for the dynamical and physical 
properties of any resulting planetary system.  

For most of its lifetime, the evolution of a protoplanetary disk is regulated 
by the interaction of gravitational and viscous stresses 
\citep[e.g.,][]{pringle81}.  Thought to be produced by magnetohydrodynamic 
turbulence \citep{balbus91}, the viscosity of the disk material acts to channel 
mass toward the star by transporting angular momentum out to larger disk radii 
\citep{lyndenbell74,hartmann98}.  This viscous evolution process decreases the 
densities and mass flow rates in the region of the disk where planets are made, 
and therefore sets a bound on the period where formation conditions are most 
favorable.  Moreover, the viscous properties of the disk control the dynamics 
of solid bodies, from the radial drift of dust particles 
\citep{weidenschilling77a} to the orbital migration of protoplanets 
\citep{lin86a}.  \citet{hartmann98} introduced some indirect methods for 
placing observational constraints on viscous evolution, based on tracking the 
decline of accretion rates with age \citep[e.g.,][]{sicilia-aguilar10} and 
characterizing disk sizes and surface density profiles with resolved millimeter 
observations \citep{kitamura02,andrews09,andrews10b,isella09,isella10}.  More 
direct approaches have focused on characterizing the anomalous viscosities, 
including attempts to measure the disk magnetic field structure 
\citep{hughes09b} and turbulent velocities \citep{hughes10b}.

Although viscous evolution controls the long-term behavior of disk material, 
there is strong observational evidence that some (if not all) disks undergo a 
faster evolutionary phase that should have important consequences for the 
assembly and development of planetary systems.  A small subsample of young 
disks exhibits a distinctive dip in their infrared spectral energy distribution 
(SED), suggesting that warm dust near the central star is substantially 
depleted compared to larger radii in the disk \citep{strom89,skrutskie90}.  
Since these so-called ``transition" disks represent only a few percent of the 
total population in nearby star-forming regions with ages up to a few Myr 
\citep{muzerolle10}, the inside-out dispersal of disk material they appear to 
trace must occur quite rapidly (a few $\times$ 0.1\,Myr, assuming all disks go 
through this phase at some point).  A variety of dissipation mechanisms have 
been proposed to explain observations of transition disks, including extensive 
particle growth \citep{tanaka05,dullemond05}, photoevaporative mass-loss in 
winds \citep{clarke01,alexander06a}, and tidal interactions with companions 
\citep[e.g.,][]{rice03,ireland08}.  As a tantalizing example of the latter 
possibility, the transition disk structures might have been modified by 
dynamical interactions with very young giant planets 
\citep[e.g.,][]{bryden99,papaloizou07}.  If that is the case, the size and 
material content of the dust-depleted region can in principle be used to 
constrain the orbit and mass of the unseen planetary companion 
\citep[e.g.,][]{lubow06,varniere06}.

Regardless of which physical mechanism is responsible, this rapid evolutionary 
phase is potentially extraordinarily important in the context of planet 
formation.  Since the short timescale for the dispersal or transformation of 
disk material can be quite influential on the accretion and migration history 
of young planets, this transitional period may play a dominant role in setting 
the final masses and orbital configurations of planetary systems.  In that 
sense, the transition disks are crucial benchmarks for planet formation models 
that rely on observational comparisons of disk properties with the demographic 
characteristics of the much older population of exoplanets.  There is an 
extensive literature on transition disk observations, ranging from their 
identification and initial probes of empirical trends 
\citep[e.g.,][]{najita07,cieza08,kim09,merin10} to detailed analyses of their 
broadband SEDs and infrared spectra 
\citep{calvet02,calvet05,dalessio05,espaillat07,espaillat10,espaillat10b}.  To 
facilitate the inclusion of this phase of disk evolution into the next 
generation of planet formation models, such studies must be coupled with firm 
observational constraints on the sizes and material contents of the 
dust-depleted zones in transition disks, as well as improved characterizations 
of their remnant mass reservoirs at larger radii.  Those goals are best 
achieved by resolving their optically thin dust emission at millimeter 
wavelengths \citep{pietu06,brown08,brown09,hughes09}.

In this paper, we present the first large sample of high angular resolution 
millimeter continuum observations of protoplanetary transition disks and 
analyze them in a homogeneous framework by constructing parametric models of 
their dust structures.  These interferometric data and their calibration are 
described in \S 2.  Detailed explanations of the adopted modeling formalism, 
radiative transfer calculations, and uncertainties in the modeling process are 
provided in \S 3.  A preliminary characterization of the transition disk 
structures is included in \S 4.  The data and models are synthesized and 
discussed in the context of disk evolution mechanisms in \S 5, with the 
key results summarized in \S 6.  A brief commentary on each individual disk is 
included in the Appendix.

\section{Observations and Data Reduction}

% TABLE 1
\begin{deluxetable*}{lcccclc}
\tablecolumns{7}
\tablewidth{0pc}
\tabletypesize{\small}
\tablecaption{SMA Observing Journal\label{obs_journal}}
\tablehead{
\colhead{Name} & \colhead{$\alpha$ [J2000]} & \colhead{$\delta$ [J2000]} & \colhead{Region} & \colhead{Array} & \colhead{UT Date} & \colhead{Reference} \\
\colhead{(1)} & \colhead{(2)} & \colhead{(3)} & \colhead{(4)} & \colhead{(5)} & \colhead{(6)} & \colhead{(7)}}
\startdata
MWC 758         & 05 30 27.53 & $+$25 19 56.9 & Tau? & E & 2008 Jan 28 & 1 \\
                &             &               &      & C & 2010 Nov 3  & 2 \\
SAO 206462     & 15 15 48.43 & $-$37 09 16.3 & Sco-OB2? & V & 2007 May 27 & 3 \\
                &             &               &      & V & 2007 Jun 8  & 3 \\
LkH$\alpha$ 330 & 03 45 48.29 & $+$32 24 11.9 & Per  & V & 2006 Nov 19 & 4 \\
                &             &               &      & C & 2010 Nov 3  & 2 \\
SR 21           & 16 27 10.28 & $-$24 19 12.8 & Oph  & V & 2007 Jun 10 & 3 \\
                &             &               &      & S & 2008 Aug 29 & 5 \\
UX Tau A        & 04 30 04.00 & $+$18 13 49.3 & Tau  & C & 2009 Oct 23 & 2 \\
                &             &               &      & E & 2010 Jan 9  & 2 \\
                &             &               &      & V & 2010 Feb 5  & 2 \\
SR 24 S         & 16 26 58.51 & $-$24 45 37.0 & Oph  & V & 2009 Mar 25 & 6 \\
                &             &               &      & C & 2009 May 4  & 6 \\
DoAr 44         & 16 31 33.46 & $-$24 27 37.4 & Oph  & C & 2008 May 13 & 5 \\
                &             &               &      & V & 2008 Apr 3  & 5 \\
LkCa 15         & 04 39 17.79 & $+$22 21 03.2 & Tau  & C & 2009 Oct 23 & 2 \\
                &             &               &      & E & 2010 Jan 13 & 2 \\
                &             &               &      & V & 2010 Feb 6  & 2 \\
RX J1615-3255   & 16 15 20.20 & $-$32 55 05.1 & Lup  & V & 2010 Feb 16 & 2 \\
                &             &               &      & C & 2010 May 1  & 2 \\
GM Aur          & 04 55 10.98 & $+$30 21 59.3 & Tau  & V & 2005 Nov 5  & 7 \\
                &             &               &      & C & 2009 Nov 6  & 8 \\
                &             &               &      & E & 2010 Jan 16 & 8 \\
DM Tau          & 04 33 48.74 & $+$18 10 09.7 & Tau  & V & 2010 Feb 17 & 2 \\
                &             &               &      & E & 2010 Sep 21 & 2 \\
                &             &               &      & C & 2010 Oct 7  & 2 \\
WSB 60          & 16 28 16.51 & $-$24 36 58.3 & Oph  & C & 2008 May 13 & 5 \\
                &             &               &      & V & 2008 Apr 3  & 5
\enddata
\tablecomments{Col.~(1): Name of host star (in order of the stellar spectral
type; see \S 3.3 and Table \ref{stars_table}).  Cols.~(2)
\& (3): Disk center coordinates (see \S 2).  Col.~(4): Star-forming region.
Col.~(5): SMA configuration; V = very extended (68-509\,m baselines), E =
extended (28-226\,m), C = compact (16-70\,m), and S = subcompact (6-70\,m).
Col.~(6): UT date of observation.  Col.~(7): Original references for SMA data:
[1] - \citet{isella10b}, [2] - this paper, [3] - \citet{brown09}, [4] -
\citet{brown08}, [5] - \citet{andrews09}, [6] - \citet{andrews10b}, [7] -
\citet{hughes09}, [8] - Hughes et al.~({\it in preparation}).}
\end{deluxetable*}

We have constructed a sample of 12 nearby transition disks with high angular 
resolution 880\,$\mu$m continuum data using new and archival observations with 
the SMA interferometer \citep{ho04}.  These targets were selected either based 
on the presence of a distinctive dip in their {\it Spitzer} InfraRed 
Spectrograph (IRS) spectrum (see the Appendix for references) and bright 
880\,$\mu$m emission ($F_{\nu} \ge 100$\,mJy) or the previous confirmation of 
a large dust-depleted cavity via direct imaging.  Those criteria introduce some 
important biases that are discussed in more detail in \S 4.  A journal of the 
SMA observations is provided in Table \ref{obs_journal}.  Since some of the 
data in this sample were previously described elsewhere 
\citep{andrews09,andrews10b,brown08,brown09,hughes09,isella10b}, details of the 
adopted instrument configurations and calibration process will not be repeated 
here.  New observations of DM Tau, LkCa 15, UX Tau, and RX J1615.3-3255 were 
obtained in 2010 February with the ``very extended" (V) configuration of the 
SMA, with baseline lengths ranging from 68 to 509\,m.  Those data were 
supplemented with shorter baseline observations in the compact (C: baseline 
lengths of 6-70\,m) and extended (E: baseline lengths of 28-226\,m, excluding 
RX J1615-3255) configurations over the past year (since 2009 October), as were 
additional observations of GM Aur (C, E), LkH$\alpha$ 330 (C), and MWC 758 (C).

For these recent observations (since fall of 2009), we used an 
expanded bandwidth capability (with the exception of the C observations of DM 
Tau) that separates each sideband into two intermediate frequency (IF) bands 
spanning 3-5\,GHz and 5-7\,GHz from the local oscillator (LO) frequency 
(340.755\,GHz, or 880\,$\mu$m).  Each IF band is composed of 24 partially 
overlapping 104\,MHz chunks, with the central chunk in the lower IF band split 
into 512 channels to sample the CO $J$=3$-$2 transition (345.796\,GHz) at a 
resolution of 0.18\,km s$^{-1}$.  All other chunks were divided into 32 coarse 
channels to measure the continuum.  Using a cycle time of $\sim$10\,minutes, 
the observing sequence interleaved disk targets with nearby quasars: J0510+180, 
3C 111, and 3C 120 for the northern disks and J1626-298, J1625-254, and 
J1604-446 for RX J1615-3255.  When the targets were at low elevations 
($<$20\degr), planets (Uranus, Mars), satellites (Titan, Callisto), and bright 
quasars (3C 273 and 3C 454.3) were observed as bandpass and absolute flux 
calibrators, depending on their availability and the array configuration.  
Weather conditions were excellent, with 225\,GHz atmospheric opacities below 
0.1 (2.0\,mm of precipitable water vapor) in all cases and substantially lower 
(down to 0.03, or 0.6\,mm of water vapor) for some observations.

The data were calibrated using the {\tt MIR} software package, treating the 
individual IF bands separately to independently assess the accuracy of the 
process.  The bandpass response was corrected based on observations of a bright 
quasar, and broadband continuum channels in each sideband and IF band were 
generated by averaging the central 82\,MHz of each chunk (except the one 
containing the CO $J$=3$-$2 transition).  The visibility amplitudes were set 
based on observations of planets or satellites, with a systematic uncertainty 
of $\sim$10\%.  The antenna-based complex gain response of the system as a 
function of time was determined with reference to the nearest local quasar 
projected on the sky, typically 3C 111, J0510+180, or J1626-298.  The other 
quasars observed in each track were used to check the quality of phase 
transfer.  Based on that information, we estimate that the millimeter ``seeing" 
generated by atmospheric phase noise and any baseline errors is small, 
$\le$0\farcs1.  The calibrated visibilities from different IF bands, sidebands, 
and array configurations were found to be in excellent agreement and combined.  
Because of the time elapsed (4-5\,yr) between observations of GM Aur and 
LkH$\alpha$ 330, the new data were corrected for proper motion 
\citep{ducourant05} before combining the visibilities.  The CO line data will 
be presented elsewhere.

% FIG 1
\begin{figure*}
\epsscale{1.2}
\plotone{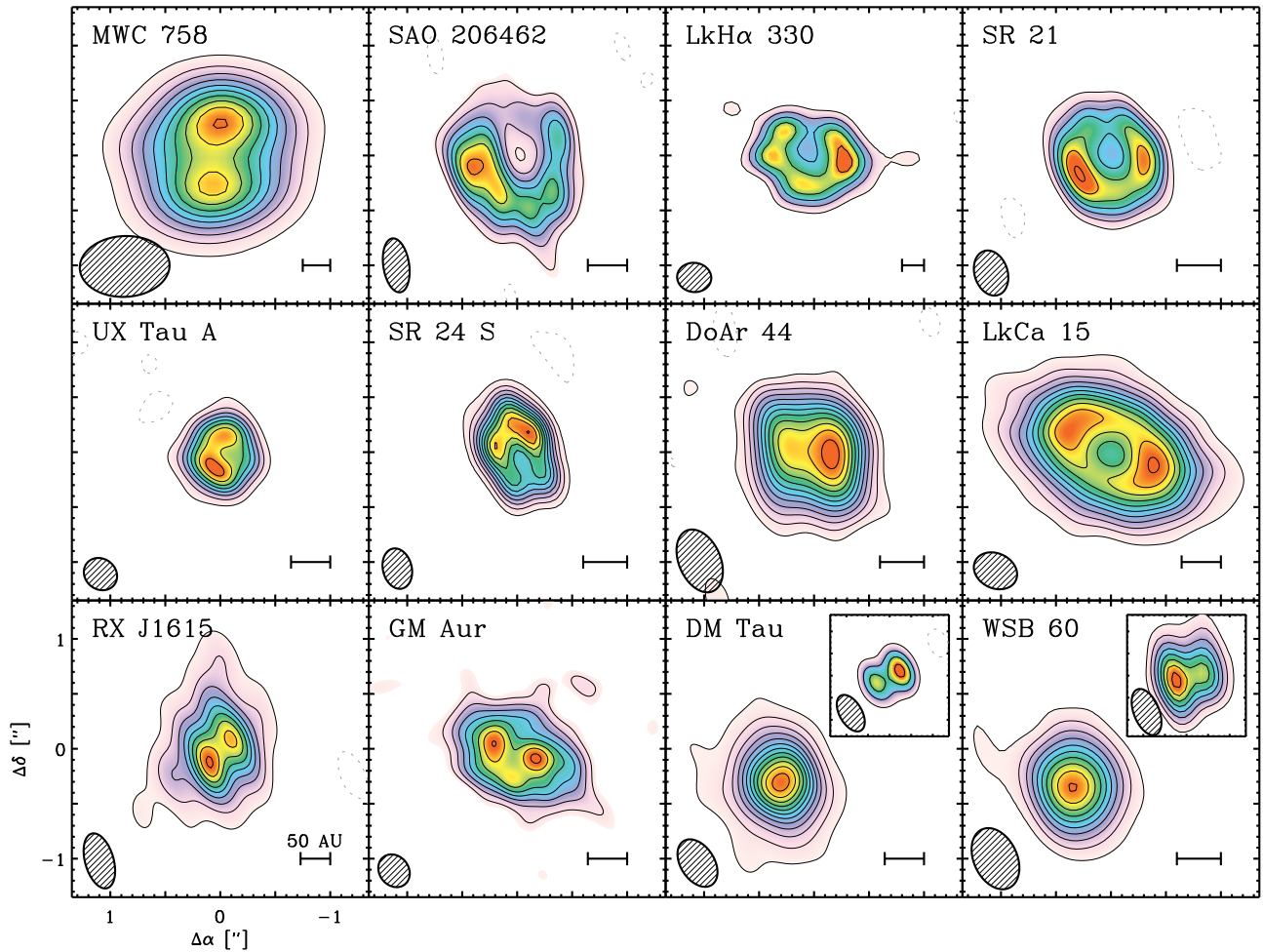}
\figcaption{SMA aperture synthesis maps of the 880\,$\mu$m continuum emission 
from this sample of transition disks.  Each panel is 2\farcs7 on a side 
(offsets are referenced to the disk centers listed in Table \ref{obs_journal}; 
see \S 2) and contains a 50\,AU projected scale bar in the lower right for 
reference.  Contours are drawn at 3\,$\sigma$ intervals, and the synthesized 
beam dimensions are marked in the lower left corner (RMS noise levels and beam 
dimensions are provided in Table \ref{image_table}).  The inset images for the 
DM Tau and WSB 60 disks were synthesized with higher angular resolution, and 
are shown to scale. \label{images}}
\end{figure*}

% TABLE 2
\begin{deluxetable}{lcccccc}
\tablecolumns{7}
\tablewidth{0pt}
\tabletypesize{\small}
\tablecaption{Aperture Synthesis Image Statistics\label{image_table}}
\tablehead{
\colhead{Name} & \colhead{$F_{\rm tot}$} & \colhead{$F_{\rm rec}$} & \colhead{$I_{\rm peak}$} & \colhead{$\sigma$} & \colhead{$\theta_b$} & \colhead{PA$_b$} \\
\colhead{} & \colhead{[Jy]} & \colhead{[Jy]} & \colhead{[mJy]} & \colhead{[mJy]} & \colhead{[\arcsec]} & \colhead{[\degr]} \\ \colhead{(1)} & \colhead{(2)} & \colhead{(3)} & \colhead{(4)} & \colhead{(5)} & \colhead{(6)} & \colhead{(7)}}
\startdata
MWC 758         & 0.18 & 0.16 & 59  & 1.1 & $0.82\times0.55$ & 93 \\
SAO 206462      & 0.62 & 0.34 & 54  & 3.5 & $0.50\times0.24$ & 9  \\
LkH$\alpha$ 330 & 0.21 & 0.21 & 28  & 2.1 & $0.31\times0.27$ & 84 \\
SR 21           & 0.40 & 0.26 & 59  & 2.6 & $0.42\times0.30$ & 18 \\
UX Tau A        & 0.15 & 0.13 & 38  & 1.5 & $0.31\times0.28$ & 52 \\
SR 24 S         & 0.55 & 0.25 & 68  & 1.8 & $0.37\times0.26$ & 14 \\
DoAr 44         & 0.21 & 0.21 & 62  & 2.7 & $0.60\times0.38$ & 23 \\
LkCa 15         & 0.41 & 0.39 & 40  & 0.9 & $0.41\times0.32$ & 64 \\
RX J1615-3255   & 0.43 & 0.30 & 76  & 2.8 & $0.52\times0.26$ & 16 \\
GM Aur          & 0.64 & 0.56 & 65  & 3.5 & $0.32\times0.27$ & 38 \\
DM Tau          & 0.21 & 0.15 & 47  & 1.3 & $0.47\times0.31$ & 33 \\
\hspace{0.25cm} ({\it inset}) & \nodata & 0.02 & 13 & 1.3 & 0.37$\times$0.21 & 31 \\
WSB 60          & 0.25 & 0.25 & 113 & 3.5 & $0.59\times0.39$ & 25 \\
\hspace{0.25cm} ({\it inset}) & \nodata & 0.20 & 67 & 3.8 & 0.44$\times$0.23 & 23
\enddata
\tablecomments{Col.~(1): Name of host star.  Col~(2): Total integrated
flux density.  Col.~(3): Recovered integrated flux density in the synthesized
maps in Figure \ref{images}.  Col.~(4): Peak flux density (in mJy
beam$^{-1}$).  Col.~(5): RMS (1\,$\sigma$) noise level in a 10\arcsec-wide
square centered on the stellar position.  Col.~(6): Synthesized beam
dimensions.  Col.~(7): Synthesized beam major axis orientation (east of
north).}
\end{deluxetable}

The combined continuum visibilities were Fourier inverted, deconvolved with the 
{\tt CLEAN} algorithm, and restored with a synthesized beam using the {\tt 
MIRIAD} software package.  Maps of the 880\,$\mu$m emission toward each target 
are exhibited together in Figure \ref{images}.  These images were synthesized 
to accentuate structures on small angular scales.  Given the variety in the 
available Fourier coverage across the sample, a range of visibility weighting 
schemes were adopted to make the maps: the typical Briggs robust parameter 
values were 0-1, although in some cases uniform (DM Tau, WSB 60) or natural 
(SAO 206462) weighting schemes were used.  The synthesized beam sizes are 
marked in the lower left of each panel and a projected 50\,AU scale bar is 
shown on the lower right.  The disks located at southern declinations have 
elongated synthesized beams because their low observing elevations prevent 
sufficient long-baseline Fourier sampling in one dimension.  The same is true 
for some northern sources that were scheduled for only a partial observing 
track (e.g., DM Tau).  The image statistics are summarized in Table 
\ref{image_table}, including the total integrated flux densities, the flux 
densities recovered in the synthesized images in Figure \ref{images} (note that 
$1-F_{\rm rec}/F_{\rm tot}$ indicates how much emission is spatially filtered 
for the adopted visibility weighting scheme), peak fluxes, noise levels, and 
beam dimensions.  Each target exhibits the continuum morphology expected for a 
resolved emission ``ring", with substantially reduced intensities near the 
stellar position.  For intermediate viewing inclinations (or elongated 
synthesized beams), these limb-brightened rings display a distinctive 
double-peaked structure on either side of the stellar position.  While that 
peak-to-peak separation offers a preliminary indication of the inner ring 
radius, more detailed modeling is required to accurately infer the disk 
structures.   

With these unique 880\,$\mu$m emission morphologies, defining a disk center 
that can be used as a reference in such modeling can be problematic.  Emission 
from the stellar photosphere is well below our detection limits at these long 
wavelengths, so there is no point-like reference in the data.  We made an 
initial estimate of the disk centers based on the optical/near-infrared stellar 
positions, corrected for any known proper motion 
\citep{perryman97,hog98,cutri03,ducourant05}.  Aside from the multiple systems 
UX Tau and SR 24 (for which proper motion estimates are particularly 
uncertain), those positions were near the apparent emission centers in the 
synthesized 880\,$\mu$m images and well within the absolute SMA astrometric 
uncertainties ($\sim$0\farcs1).  Those positions were then refined using a grid 
search that minimized the imaginary component of the visibilities (i.e., the 
position for which the RMS of the deviations from zero imaginary flux was 
smallest).  In principle, the imaginary component of a symmetric source should 
be zero.  Since our models will assume axisymmetric structures, this assumption 
for centering seems appropriate.  An {\it a posteriori} comparison of the data 
and our modeling results (see \S 4) confirms that there is no statistically 
significant evidence for asymmetric emission structures on the angular scales 
probed by the SMA.  The adopted reference centers are listed in Table 
\ref{obs_journal}.  

% FIG 2
\begin{figure*}
\epsscale{1.10}
\plottwo{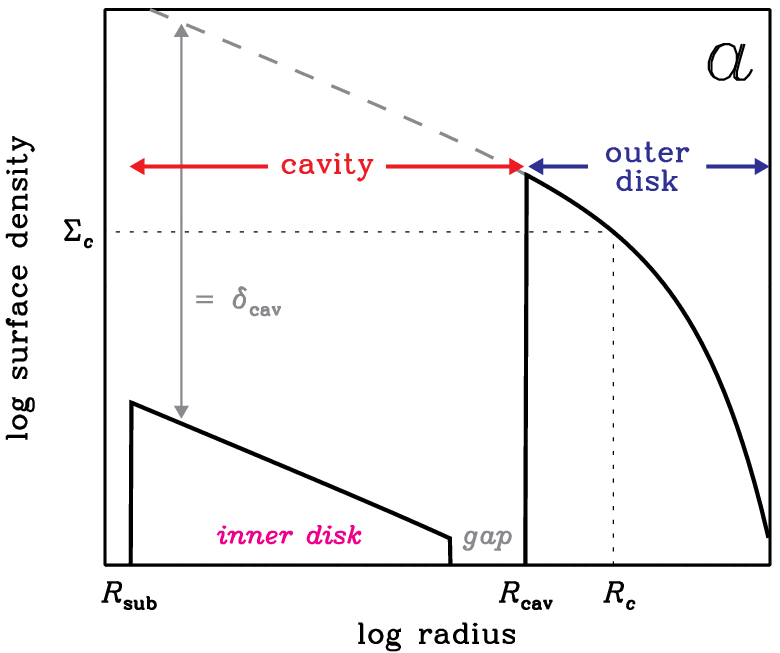}{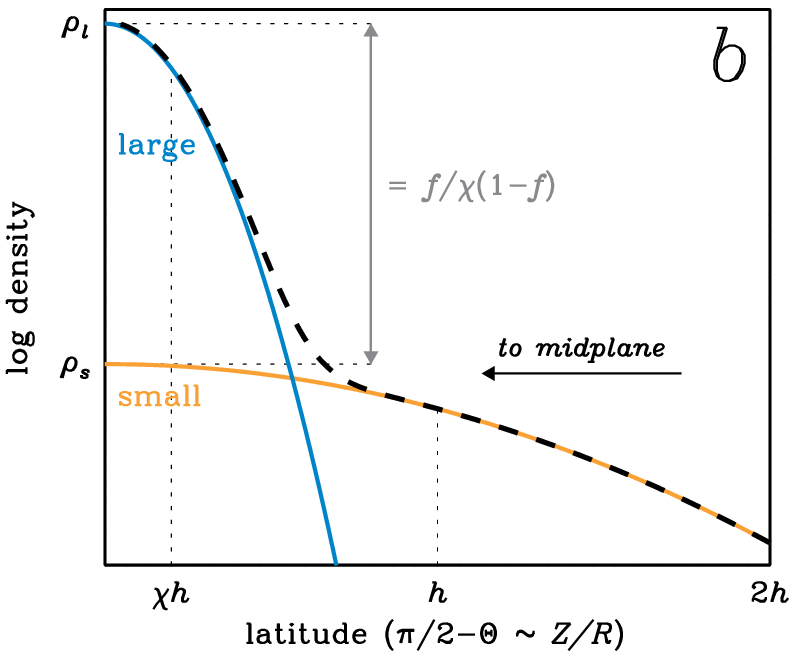}
\figcaption{($a$) A generic surface density model.  The outer disk ($R > R_{\rm
cav}$) follows the global $\Sigma_g$ profile, overlaid as a dashed curve.  The
surface densities in the cavity ($R \le R_{\rm cav}$) are scaled down by a
factor $\delta_{\rm cav}$.  The dust-depleted inner disk is truncated at the
sublimation radius ($R_{\rm sub}$) and extends outward to an arbitrary fixed
radius (10\,AU).  The gap between 10\,AU and $R_{\rm cav}$ is assumed to be
empty.  ($b$) A generic vertical density profile.  The blue curve tracks the
``large" dust population, concentrated toward the midplane with a scale height
$\chi h$ (in the models used here, $\chi$ is fixed to 0.2).  The orange curve
marks the ``small" dust population that dominates in the disk atmosphere, and
the dashed curve traces the composite profile.  The fraction of the column
density contributed by the large dust population is denoted $f$ (in the models
used here, $f$ is fixed to 0.85).  \label{sigma_demo}}
\end{figure*}

\section{Models of Transition Disk Structures}

The physical structures of transition disks are quite complex, and the 
few studies aimed at their characterization using spatially resolved data have 
adopted very different approaches \citep{pietu06,hughes07,hughes09,brown08,brown09,andrews09,andrews10a,andrews10b,isella10,isella10b}.  In this study, our 
goal is to analyze this large sample of transition disks in a homogeneous 
framework, emphasizing the new constraints on basic structural parameters that 
are responsible for the key features of the high angular resolution SMA data.  
These initial results should help guide more in-depth, individualized studies 
in the future.  The modeling performed here is based on the previous work 
described by \citet{andrews09,andrews10a,andrews10b} and \citet{hughes10}, but 
includes some modifications meant to imitate (parametrically) the detailed 
treatments of \citet{calvet02,calvet05}, \citet{dalessio05}, and 
\citet{espaillat07,espaillat10}.  In the following sections, we describe a 
parametric model of the disk structure (\S 3.1), our assumptions about the 
dust populations (\S 3.2) and stellar properties (\S 3.3), and the radiative 
transfer calculations and our approach for comparing the models with 
observations (\S 3.4).  Finally, we discuss the observable effects of the model 
parameters and important uncertainties (\S 3.5), and provide a concise summary 
of some key issues related to the modeling (\S 3.6) for those less interested 
in the details.

\subsection{Parametric Density Model}

The model structure includes two distinct radial zones: a dust-depleted 
``cavity" ($R \le R_{\rm cav}$) and an ``outer disk" reservoir ($R > R_{\rm 
cav}$).  We define a global surface density profile, appropriate for a simple 
accretion disk with a power-law radial distribution of time-independent 
viscosities \citep[$\nu \propto R^{\gamma}$; see][]{lyndenbell74,hartmann98},
\begin{equation}
\Sigma_g = \Sigma_c \left( \frac{R}{R_c} \right)^{-\gamma} \exp \left[ - \left( \frac{R}{R_c} \right)^{2-\gamma} \right]
\end{equation}
where $R_c$ is a characteristic scaling radius.  This global profile is applied 
to the material in the outer disk: $\Sigma(R > R_{\rm cav}) = \Sigma_g$.  The 
densities in the cavity are scaled down by a constant depletion factor, 
$\delta_{\rm cav}$, such that $\Sigma(R\le R_{\rm cav}) = \delta_{\rm cav} 
\Sigma_g$.  To facilitate a more homogeneous comparison of $\delta_{\rm cav}$ 
values for a sample with a range of cavity sizes ($R_{\rm cav}$), we decomposed 
the cavity into a depleted ``inner disk" of fixed extent (out to $R = 10$\,AU) 
and an {\it empty} ``gap" between 10\,AU and $R_{\rm cav}$ (the arbitrary 
choice of the inner disk extent is discussed in \S 3.5).  Unless noted 
otherwise, the inner disk is truncated at the dust sublimation 
radius, $R_{\rm sub} \approx 0.07 (L_{\ast}/$L$_{\odot})^{1/2}$\,AU 
\citep[assuming $T_{\rm sub} = 1500$\,K;][]{dullemond01}.  Figure 
\ref{sigma_demo}a shows a generic surface density model for reference.  

The vertical distribution of dust has been modified from our previous models, 
in an effort to more realistically treat the irradiation heating of the disk 
atmosphere.  We have introduced a vertical gradient in the dust size 
distribution, such that larger grains are concentrated toward the midplane and 
smaller grains are distributed to larger vertical heights.  In practice, this 
parameterization crudely mimics dust sedimentation 
\citep[e.g.,][]{dullemond04b,dalessio06}.  At each radius, a population of 
``small" grains represents a small fraction of the total column density, $(1-f) 
\Sigma$, and is distributed vertically like a Gaussian with scale height $h = 
h_c (R/R_c)^{\psi}$.  Conversely, most of the total column ($f\Sigma$) is 
composed of a ``large" grain population, condensed toward the midplane with a 
scale height $\chi h$ (where $\chi \le 1$).  In a spherical coordinate system 
with azimuthal and mirror symmetry, the two-dimensional density structures for 
each dust population are 
\begin{equation}
\rho_s = \frac{(1-f)\Sigma}{\sqrt{2\pi} R h} \exp \left[- \frac{1}{2} \left( \frac{\pi/2 - \Theta}{h} \right) ^2 \right]
\end{equation}
\begin{equation}
\rho_l = \frac{f\Sigma}{\sqrt{2\pi} R \chi h} \exp \left[- \frac{1}{2} \left( \frac{\pi/2 - \Theta}{\chi h} \right) ^2 \right]
\end{equation}
where the subscripts `$s$' and `$l$' denote the small and large dust 
populations, respectively, $\Theta$ is the (vertical) latitude coordinate 
measured from the pole ($\Theta = 0$) to the equator (the midplane, $\Theta = 
\pi/2$), and the value of $\Sigma$ depends on the radial location as outlined 
above.  Note that in this coordinate system, the scale height $h$ is an angle: 
the scale height in physical (distance) units is $H \approx hR$.  A generic 
vertical density profile with this parametric form is shown in Figure 
\ref{sigma_demo}b.

The distinctive infrared spectra of transition disks require small, but 
important, local modifications to the vertical structure near the disk edges at 
$R_{\rm sub}$ and $R_{\rm cav}$ \citep{natta01,calvet02,espaillat10}.  At those 
locations, a large surface area of material is exposed directly (or with 
relatively low attenuation; see \S 3.5) to stellar irradiation.  That exposure 
substantially heats a thin layer of material and ``puffs" up the local dust 
distribution \citep{dullemond04a,dalessio05}.  To mimic the emission signatures 
of these structures $-$ referred to as the inner ``rim" (at $R_{\rm sub}$) and 
cavity ``wall" (at $R_{\rm cav}$) $-$ we permit the local value of $h$ to be 
scaled up parametrically (to $h_{\rm rim}$ and $h_{\rm wall}$, respectively).  
Those perturbations are joined to the global scale height distribution 
exponentially over a small radial width (fixed here to $\Delta R = 0.1$\,AU).  

This structure model has 11 parameters: 5 describe the surface density profile, 
\{$\Sigma_c$, $\gamma$, $R_c$, $\delta_{\rm cav}$, $R_{\rm cav}$\}, and 6 
others characterize the vertical distribution of dust, \{$h_c$, $\psi$, $\chi$, 
$f$, $h_{\rm rim}$, $h_{\rm wall}$\}.  Some of these parameters can not be 
constrained with the available data, and were therefore fixed to representative 
values.  The surface density gradient was kept at $\gamma = 1$, in line with 
the results for ``normal" disks with continuous dust distributions 
\citep{andrews09,andrews10b}.  We set the settling parameters to $\chi = 0.2$ 
(the large grains are distributed to 20\%\ of the scale height, $h$) and $f = 
0.85$ (85\%\ of the total column is composed of the large grain population).  
The remaining parameters are allowed to freely vary, although substantial 
degeneracies remain (see \S 3.5).

\subsection{Dust Populations}

\begin{figure}
\epsscale{1.1}
\plotone{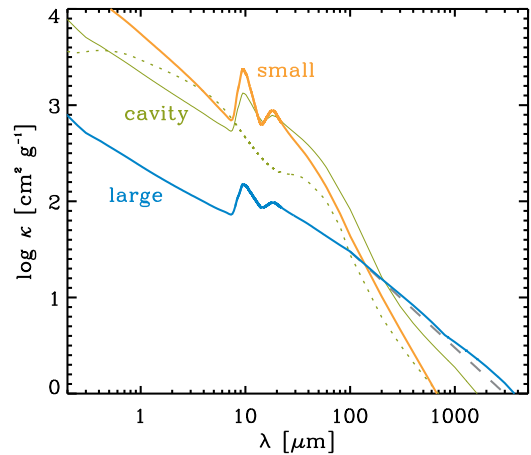}
\figcaption{The opacity spectra for representative dust populations.  The
opacities for the small (orange) and large (blue) dust populations in the outer
disk are fixed; the latter is similar to the canonical \citet{beckwith90}
values (gray dashed line).  As illustrative examples, the opacities for two
different dust populations we have used inside the disk cavities are also shown
(green): \{$s_{\rm max}$, $p$\} = \{10\,$\mu$m, 3.5\} (solid) and \{1\,$\mu$m,
2.5\}(dotted).  \label{opacities}}
\end{figure}

A parametric density structure must be populated with dust grains of a 
given size distribution and composition.  For the sake of homogeneity, we 
assume dust grains with mineral abundances as in the interstellar medium (ISM) 
\citep[see][]{weingartner01}, despite some evidence in individual cases for 
different compositions.  Each dust population has a power-law distribution of 
sizes ($s$), $n(s) \propto s^{-p}$, from $s_{\rm min} = 0.005$\,$\mu$m to a 
specified $s_{\rm max}$.  The dust properties in the outer disk are fixed: 
``small" grains have $s_{\rm max} = 1$\,$\mu$m, ``large" grains have $s_{\rm 
max} = 1$\,mm, and $p = 3.5$.  The opacity spectrum for a given size 
distribution was derived from Mie calculations.  Since most of the mass is in 
``large" grains, the selection of $s_{\rm max} \sim \lambda = 1$\,mm is 
significant because it tends to maximize the millimeter-wave dust opacity.  
Therefore, the densities inferred from the optically thin emission we have 
observed with the SMA can be considered lower bounds if much larger particles 
are present \citep{dalessio06,draine06}.  The 880\,$\mu$m dust opacity for the 
large grain population is 3.6\,cm$^2$ g$^{-1}$.  The dust size distribution in 
the inner disk, inner rim, and cavity wall are assumed to be identical, but are 
characterized by free parameters for their size distribution, \{$s_{\rm max}$, 
$p$\} (see \S 3.4 regarding the selection of these parameter values).  Some 
representative opacity spectra are shown in Figure \ref{opacities}.  

% FIG 3

\subsection{Stellar Properties}

Passive irradiation is the only heating mechanism considered here, and 
therefore the properties of the central star are pivotal in setting the thermal 
structure of a model.  Careful selection of an input stellar spectrum is 
particularly important for modeling transition disks, as the infrared 
disk-to-star contrast is often low.  Therefore, the details of the stellar 
spectrum can have a pronounced impact on how the small amount of dust inside 
the cavity is interpreted.  To derive stellar input spectra, we first assigned 
effective temperature ($T_{\rm eff}$) values based on spectral classifications 
in the literature, using the conversions advocated by \citet{kenyon95} and 
\citet{luhman99a} for types earlier or later than M0, respectively.  
Luminosities ($L_{\ast}$) were determined by matching reddened, scaled spectral 
synthesis models \citep[][for a fixed $T_{\rm eff}$]{lejeune97} to the 
optical/near-infrared photometry (see the Appendix).  We utilized the 
extinction curve derived by \citet{mcclure09}, which at these wavelengths is 
identical to the \citet{mathis90} law for a large total-to-selective extinction 
value, $R_V = 5$.  For cases without dynamical mass constraints, this process 
was iterated for different stellar masses ($M_{\ast}$) estimated from the 
\citet{siess00} pre-main sequence models.  The stellar parameters used to 
generate input spectra are compiled in Table \ref{stars_table}.

% TABLE 3
\begin{deluxetable*}{lcccccccccl}
\tablecolumns{11}
\tablewidth{0pt}
\tabletypesize{\small}
\tablecaption{Stellar and Accretion Properties\label{stars_table}}
\tablehead{
\colhead{Name} & \colhead{SpT} & \colhead{$T_{\rm eff}$} & \colhead{$A_V$} & \colhead{$d$} & \colhead{$L_{\ast}$} & \colhead{$R_{\ast}$} & \colhead{$M_{\ast}$} & \colhead{$\dot{M}$} & \colhead{$\log{L_X}$} & \colhead{Ref} \\
\colhead{} & \colhead{} & \colhead{[K]} & \colhead{} & \colhead{[pc]} & \colhead{[L$_{\odot}$]} & \colhead{[R$_{\odot}$]} & \colhead{[M$_{\odot}$]} & \colhead{[M$_{\odot}$ yr$^{-1}$]} & \colhead{[erg s$^{-1}$]} & \colhead{} \\
\colhead{(1)} & \colhead{(2)} & \colhead{(3)} & \colhead{(4)} & \colhead{(5)} & \colhead{(6)} & \colhead{(7)} & \colhead{(8)} & \colhead{(9)} & \colhead{(10)} & \colhead{(11)}}
\startdata
MWC 758         & A8 & 7580 & 0.0 & 200 & 15  & 2.25 & 1.8  & $1\times10^{-8}$  & \nodata  & 1, 2, 3, \nodata \\
SAO 206462      & F4 & 6590 & 0.3 & 140 & 7.8 & 2.15 & 1.6  & $6\times10^{-9}$  & 29.7     & 4, 5, 6, 6       \\
LkH$\alpha$ 330 & G3 & 5830 & 3.1 & 250 & 15  & 3.75 & 2.2  & $2\times10^{-9}$  & \nodata  & 7, 5, 8, \nodata \\
SR 21           & G3 & 5830 & 6.3 & 125 & 10  & 3.15 & 2.0  & $<1\times10^{-9}$ & 30.0     & 9, 5, 10, 11     \\
UX Tau A        & G8 & 5520 & 1.9 & 140 & 3.5 & 2.05 & 1.5  & $1\times10^{-8}$  & 30.3     & 12, 5, 12, 13    \\
SR 24 S         & K2 & 4990 & 7.0 & 125 & 4.0 & 2.70 & 2.0  & $1\times10^{-8}$  & 30.0     & 14, 5, 10, 15    \\
DoAr 44         & K3 & 4730 & 2.2 & 125 & 1.4 & 1.75 & 1.3  & $9\times10^{-9}$  & 29.9     & 12, 5, 12, 16    \\
LkCa 15         & K3 & 4730 & 1.7 & 140 & 1.2 & 1.65 & 1.01 & $2\times10^{-9}$  & $< 29.6$ & 12, 17, 18, 19   \\
RX J1615-3255   & K5 & 4350 & 0.4 & 185 & 1.3 & 2.00 & 1.1  & $4\times10^{-10}$ & 30.4     & 20, 5, 5, 21     \\
GM Aur          & K5 & 4278 & 1.1 & 140 & 1.0 & 1.85 & 0.84 & $1\times10^{-8}$  & $< 29.7$ & 22, 23, 18, 19   \\
DM Tau          & M1 & 3705 & 0.6 & 140 & 0.3 & 1.25 & 0.53 & $6\times10^{-9}$  & $< 29.7$ & 24, 17, 18, 25   \\
WSB 60          & M4 & 3270 & 3.7 & 125 & 0.2 & 1.40 & 0.25 & $2\times10^{-9}$  & \nodata  & 26, 5, 10, \nodata
\enddata
\tablecomments{Col.~(1): Name of host star.  Col.~(2): Spectral type.
Col.~(3): Visual extinction.  Col.~(4): Effective temperature.  Col.~(5):
Luminosity.  Col.~(6): Estimated distance to star-forming region.  Col.~(7):
Stellar radius.  Col.~(8): Stellar mass, determined either directly from CO
spectral images or indirectly from pre-main sequence evolution models (see \S
3.2).  Col.~(9): Accretion rate.  Col.~(10): X-ray luminosity or
3\,$\sigma$ upper limit.  Note that the value for UX Tau A may include
contributions from the additional stars in the multiple system.  To our
knowledge, no X-ray measurements of WSB 60, LkH$\alpha$ 330, or MWC 758 have
been published to date.  Col.~(11): Literature references for spectral
classifications, $M_{\ast}$, $\dot{M}$, and $L_X$: [1] - \citet{beskrovnaya99},
[2] - \citet{chapillon08}, [3] - \citet{eisner09}, [4] - \citet{dunkin97}, [5]
- this paper (see \S 3.3), [6] - \citet{grady09}, [7] - \citet{coku79}, [8] -
\citet{salyk09}, [9] - \citet{prato03}, [10] - \citet{natta06}, [11] -
\citet{grosso00}, [12] - \citet{espaillat10}, [13] - \citet{koenig01}, [14] -
\citet{luhman99}, [15] - \citet{pillitteri10}, [16] - \citet{montmerle83}, [17]
- \citet{pietu07}, [18] - \citet{ingleby09}, [19] - \citet{neuhauser95}, [20] -
\citet{wichmann99}, [21] - \citet{krautter97}, [22] - \citet{white01}, [23] -
\citet{dutrey98}, [24] - \citet{kenyon95}, [25] - \citet{damiani95}, [26] -
\citet{wilking05}.}
\end{deluxetable*}

\subsection{Radiative Transfer and Modeling Procedure}

For a given set of parameters, a model structure was defined on a $200 \times 
90$-cell grid in spherical coordinates, \{$R$, $\Theta$\}.  The radial grid was 
logarithmically spaced from the sublimation radius to 1500\,AU, with local 
refinements near $R_{\rm sub}$, 10\,AU (the fixed edge of the inner disk), and 
$R_{\rm cav}$ to sample the optical depth gradients at those locations with 
higher resolution.  The latitude grid was linear, with three distinct 
resolution levels that depend on the vertical structure of the model: high near 
the condensed midplane, coarse at high latitudes near the pole, and 
intermediate in the disk atmosphere.  Simulations of the stellar irradiation of 
that model density grid were conducted to determine the dust temperature 
structure, using the two-dimensional axisymmetric Monte Carlo radiative 
transfer code {\tt RADMC} \citep[see][]{dullemond04a}.  Those calculations 
explicitly accounted for the finite size of the star ($R_{\ast}$; see Table 
\ref{stars_table}) to properly treat any shadowing effects from the inner rim 
or cavity wall (stellar limb darkening is not considered).  For a specified 
viewing geometry (a disk inclination, $i$, and major axis orientation, PA), a 
raytracing algorithm was used on the results of the radiative transfer 
calculations to compute a synthetic SED and set of 880\,$\mu$m continuum 
visibilities (sampled at the same spatial frequencies observed with the SMA).

Aided by a series of refined grid searches in parameter-space, this initial 
modeling effort was primarily conducted manually.  A confirmation of fit 
quality was performed by comparing the synthetic data for a given parameter set 
with the observed SED and the deprojected, elliptically-averaged 880\,$\mu$m 
visibility profile \citep[for details, see][]{andrews09}.  For this model 
setup and such a large sample, parameter estimation with automated minimization 
algorithms would be computationally prohibitive: the free parameter-space is 
too large (8 density parameters, 2 dust parameters, and 2 viewing geometry 
parameters) and degenerate.  Moreover, there are non-trivial technical 
obstacles: for example, great care is required in defining the model grid for a 
given parameter set, to ensure proper treatment of large optical depth 
gradients and small-scale features like the rim and wall \citep[see 
also][]{andrews09}.  Regardless of these challenges, good (albeit not unique, 
best-fit) matches to the data were found with a simple and systematic approach.

In most cases, the viewing geometry was fixed based on models of CO spectral 
images in the literature \citep{dutrey98,pietu07,isella10b,lyo11} or the new 
$J$=3$-$2 data from the SMA (for LkH$\alpha$ 330, UX Tau A, and RX 
J1615-3255; details will be presented elsewhere).  No such information is 
available for the Ophiuchus disks (SR 21, SR 24 S, DoAr 44, and WSB 60): 
estimates of \{$i$, PA\} in those cases were inferred from the aspect ratio and 
orientation of the continuum, and are considerably uncertain 
\citep[see][]{andrews09,andrews10b}.  In practice, the cavity size, $R_{\rm 
cav}$, was determined from the null position in the SMA visibility profile (see 
\S 3.5).  The dust parameters in the cavity, \{$s_{\rm max}$, $p$\}, were set 
based on the strength of the 10\,$\mu$m silicate feature, and the density 
contrast ($\delta_{\rm cav}$) and rim height ($h_{\rm rim}$) were scaled to 
match the weak infrared excess from the inner disk.  The height of the cavity 
wall ($h_{\rm wall}$) was adjusted to reproduce the shape and rise of the 
mid-infrared spectrum, and the scale height parameters, \{$h_c$, $\psi$\}, were 
modified based on the far-infrared SED.  The characteristic radius, $R_c$, and 
density normalization, $\Sigma_c$, were inferred from the shape and amplitude 
of the 880\,$\mu$m visibility profile.

\subsection{Parameter Effects and Degeneracies}

\begin{figure}
\epsscale{1.1}
\plotone{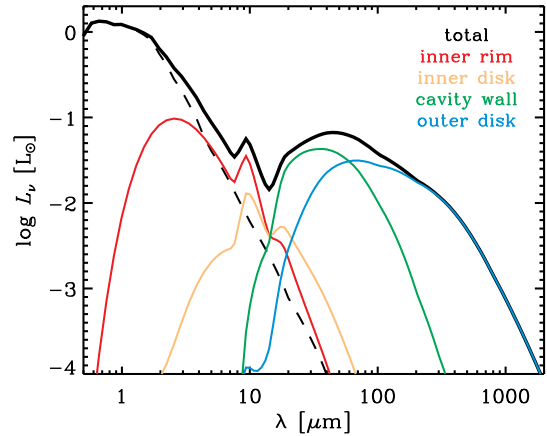}
\figcaption{A fiducial model SED decomposed into individual structural
elements.  The ordinates are luminosity densities, $L_{\nu} = 4 \pi d^2 \nu
F_{\nu}$, in solar units.  The model assumes a K3 star ($\sim$1\,M$_{\odot}$,
2\,R$_{\odot}$) irradiates a disk with a total mass 0.07\,M$_{\odot}$, $R_c =
75$\,AU, and scale height distribution with $H_{\rm 100} = 3$\,AU and $\psi =
0.2$.  The densities inside $R_{\rm cav} = 40$\,AU have been scaled down by a
factor $\delta_{\rm cav} \approx 10^{-6}$, and the inner rim and wall have been
artificially puffed up a factor of $\sim$3 over their local $H$ value ($H_{\rm
sub} \approx 0.005$\,AU, or 0.5\,$R_{\ast}$, and $H_{\rm wall} \approx 3$\,AU,
respectively).  The dust in the inner disk is the same as in the outer disk
atmosphere, with $s_{\rm max} = 1$\,$\mu$m and $p = 3.5$.  The solid black
curve is the sum of the colored curves and the contribution of the stellar
photosphere (dashed).  \label{SEDzones}}
\end{figure}

\begin{figure*}
\epsscale{1.1}
\plotone{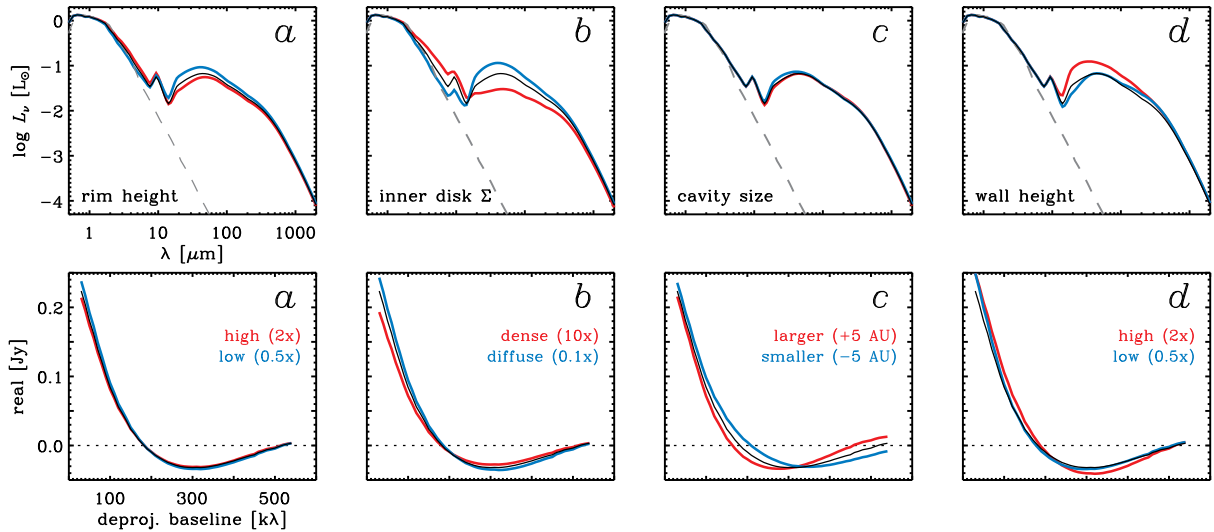}
\figcaption{Effects of small modifications to key transition disk parameters on
the SED ({\it top}) and 880\,$\mu$m visibility profile ({\it bottom}).  The
black curve is the fiducial model shown in Figure \ref{SEDzones}.  The effects
of parameter increases relative to the fiducial model are shown in red,
decreases in blue: ($a$) height of the inner rim ($h_{\rm rim}$), a factor of
2; ($b$) inner disk density ($1/\delta_{\rm cav}$), a factor of 10; ($c$)
cavity size ($R_{\rm cav}$), $\pm$5\,AU; and ($d$) cavity wall height ($h_{\rm
wall}$), a factor of 2. \label{pareffects}}
\end{figure*}

To reinforce the intuition used to model these data and provide an overview of 
the individual parameter effects, it is helpful to decompose the disk structure 
into four essential elements: the inner rim, dust-depleted inner disk, cavity 
wall, and outer disk.  Each element has a distinctive impact on the dust 
emission spectrum, and together they are capable of producing a remarkable 
diversity of observational signatures.  To illustrate those signatures, we 
define a fiducial reference model and show separately the contributions from 
each structural element on the SED in Figure \ref{SEDzones} (fiducial 
parameters are listed in the caption).  Figure \ref{pareffects} explores how 
the SED and 880\,$\mu$m visibility profile respond to modest changes in the 
parameters of specific interest for transition disks: the rim and wall heights 
($h_{\rm rim}$ and $h_{\rm wall}$, respectively), the inner disk mass (set by 
$\delta_{\rm cav}$), and the cavity size ($R_{\rm cav}$).  The effects of more 
global structure parameters were discussed by \citet{andrews09}.  

The emission from the directly-illuminated dust in the inner rim dominates the 
near-infrared SED (see Figure 4).  The luminosity of that hot component is set 
by the size (and optical depth) of the rim, such that a more ``puffed-up" rim 
(larger $h_{\rm rim}$) produces a brighter excess.  A sufficiently high (and 
dense) rim can intercept enough starlight to shadow material at larger radii.  
The attenuated irradiation field that reaches those components produces less 
heating and therefore a fainter infrared spectrum (see Figure 
\ref{pareffects}a).  While this shadowing effect has been explored for normal 
disks \citep{dullemond00,dullemond01}, its effects can be pronounced for 
transition disks because much of their infrared spectrum is produced by the 
directly-illuminated portion of the cavity wall \citep{espaillat10}.  
\citet{espaillat10b} have shown that the {\it Spitzer} IRS spectra for many 
transition disks exhibit remarkable ``see-saw" variability, such that an 
increase in the near-infrared emission is correlated with a decrease at 
mid-infrared wavelengths (and vice versa).  They make a compelling argument 
that this variability is produced by changes in the height of the inner rim, as 
demonstrated in Figure \ref{pareffects}a \citep[see 
also][]{dullemond03,flaherty10}.  

% FIG 4

% FIG 5

Beyond the rim, the tenuous inner disk emits an infrared spectrum corresponding 
to the temperature range between $R_{\rm sub}$ and 10\,AU.  In cases where this 
material is optically thin, the luminosity scales directly with the density 
(parameterized here by $\delta_{\rm cav}$).  However, small amounts of dust in 
the cavity can be optically thick.  Even for substantial optical depths, the 
infrared luminosity still partly scales with density: for a given $h$, a denser 
disk has more dust at larger heights above the midplane, increasing the height 
and temperature of the infrared photosphere.  The result is a brighter (and 
hotter) excess spectrum.  The same is true for the inner rim, and therefore 
$\delta_{\rm cav}$ can also have an impact on shadowing.  Provided a 
sufficiently high rim height, $h_{\rm rim}$ and $\delta_{\rm cav}$ are 
degenerate (see Figure \ref{pareffects}a,b): higher rims can compensate for 
lower densities (at least partially).  Moreover, the inner disk size (fixed 
here to 10\,AU) is also degenerate with $\delta_{\rm cav}$: a more extended 
inner disk (a smaller gap, or even no gap) can usually be accomodated by 
decreasing $\delta_{\rm cav}$.  Other fixed assumptions about the inner disk 
structure can also have comparable effects (e.g., the width of the rim, shape 
of the density profile, local grain properties, etc.).  The important point to 
be made is that even a small amount of ($\sim$$\mu$m-sized) dust inside the 
cavity can have a substantial impact on the SED, depending rather sensitively 
on its detailed spatial distribution and grain properties.  Therefore, without 
spatially resolved observations, the detailed characteristics of the material 
inside transition disk cavities are strongly model-dependent and highly 
uncertain.

Despite our ignorance of the detailed structures of the inner disk and rim, the 
cavity size can be measured from the SMA data with little ambiguity.  Past 
efforts to indirectly infer $R_{\rm cav}$ relied on estimating the wall 
temperature from the shape of the mid-infrared spectrum 
\citep[e.g.,][]{dalessio05,calvet05,espaillat07}.  Those models have become 
increasingly sophisticated to try and account for the irradiation gradients 
caused by shadowing \citep{espaillat10}, but still suffer from the implicit 
uncertainties in the inner disk properties (not to mention the detailed 
characteristics of the wall itself).  However, a {\it direct} measurement of 
$R_{\rm cav}$ is possible with a resolved, optically thin tracer like the 
millimeter continuum.  The sharp emission contrast between the faint dust in 
the cavity and the bright material in the outer disk produces a clear ``ring" 
morphology at 880\,$\mu$m.  The Fourier transform of that emission profile 
(i.e., the visibility profile) has distinctive nulls at discrete spatial 
frequencies, such that the location of the first null (where the real part of 
the visibilities change sign) is a relatively unambiguous measure of the inner 
edge of the ring \citep[$R_{\rm cav}$; see][]{hughes07}.  Figure 
\ref{pareffects}c demonstrates the sensitivity of this null position to $R_{\rm 
cav}$; modifications on the order of 10\%\ are readily detectable for typical 
noise levels.  However, we should note that these null positions also have a 
roughly linear dependance on the radial gradient of the surface brightness 
profile, which is set by the product of $\Sigma$ and the midplane temperature 
profile.  Since the latter is approximately a power-law in these models 
($T_{\rm mid} \propto R^{-q}$; typically $q \approx 0.5$), the $R_{\rm cav}$ 
value inferred from a fixed null position is proportional to the sum $\gamma + 
q$.  Since $\gamma = 1$ is fixed in these models, we under-estimate the 
uncertainty on $R_{\rm cav}$: allowing $\gamma$ to vary by $\pm0.5$ would 
contribute an additional $\sim$10-15\%\ uncertainty to the cavity sizes.

In a simple model with an empty cavity, the directly-irradiated wall emits a 
thermal spectrum corresponding to the equilibrium temperature at its location 
($R_{\rm cav}$) with a luminosity proportional to its vertical extent 
\citep{calvet02,dalessio05}.  To first order, the behavior expected from such a 
model is qualitatively reproduced in Figure \ref{pareffects}d: smaller cavities 
have hotter walls and narrower infrared SED dips.  As mentioned above, there 
are slight deviations in the details because shadowing can induce vertical 
temperature gradients in the wall.  In principle, a high wall can shadow the 
outer disk if the dust is sufficiently settled (low $h_c$ and/or $\psi$).  The 
detailed properties of the cavity wall (e.g., shape, thickness) are not well 
constrained by the SED alone; no current observations have sufficient 
resolution to isolate such a (presumably) small feature.  

Although all disk modeling suffers from degeneracies between structure 
parameters and dust properties, it is important to explicitly highlight one key 
issue in these transition disk models.  The standard mass-opacity degeneracy 
for optically thin emission \citep[e.g.,][]{beckwith91,aw05} and the intricate 
relationship between irradiation, heating, and grain properties in the disk 
atmosphere \citep{dalessio99,dalessio01,dalessio06,chiang01} are important, but 
will not be reviewed again here.  The degeneracy of interest is related to how 
our method of inferring densities in the inner disk (i.e., $\delta_{\rm cav}$) 
depends on the dust opacities inside the cavity.  Since we lack any resolved 
millimeter signal inside $R_{\rm cav}$ (although see \S 4 regarding LkCa 15), 
the inner disk densities are determined solely from the infrared 
spectrum.  In most cases, the relatively strong silicate emission features we 
observe suggest that the cavity is populated by ``small" grains (i.e., $s_{\rm 
max} \approx 1$-10\,$\mu$m or less).  To be conservative, we do not assume that 
any additional material is present.  This small dust becomes optically thick 
even at the relatively low column densities ($\sim$10$^{-4}$\,g cm$^{-2}$, give 
or take an order of magnitude; see Figure \ref{opacities}) that are present in 
some non-negligible part of the inner disk.  The sizes of these high optical 
depth regions are effectively responsible for setting $\delta_{\rm cav}$: for a 
given structure, more mass would over-produce the infrared luminosities.  In 
some sense, this aspect of the modeling tends to minimize $\delta_{\rm cav}$ 
(or maximize the depletion of dust densities inside $R_{\rm cav}$).  However, 
more material could easily be accomodated in the inner disk if it has lower 
infrared opacities (e.g., $s_{\rm max}$ larger than a few $\mu$m; see \S 4).  

Finally, a few notes about the outer disk parameters are in order \citep[for a 
more complete discussion, see][]{andrews09}.  It is important to emphasize 
that the nature of transition disk structures makes it difficult to constrain 
radial gradient parameters ($\gamma$ or $\psi$).  For smaller disks, the same 
is true for the characteristic radius, $R_c$.  The issue is limited dynamic 
range; only a few radial resolution elements can probe the millimeter emission 
in a ring (which is why we chose to fix $\gamma$), and the far-infrared SED 
that is less affected by the cavity is only sparsely sampled.  While we have a 
reasonable constraint on the vertical distribution of dust ($h_c$) from the 
infrared luminosities, the current lack of far-infrared data means that we do 
not have a good handle on the scale height gradient, $\psi$, for these 
transition disks with large cavities.  Linking these models with scattered 
light observations that specialize in constraining the flaring angle (i.e., 
$\psi$) will be beneficial in the future.  

Although they were fixed, the 
settling parameters \{$\chi$, $f$\} merit a brief discussion.  The fraction of 
large dust grains, $f$, helps set the height where stellar photons are 
absorbed: smaller $f$ means an increased optical depth (more abundant small 
grains) at a fixed height, and therefore a slightly warmer atmosphere that 
emits more infrared radiation.  The effects of $\chi$, the fractional scale 
height of the large grain population, are subtle.  In practice, $\chi$ modifies 
the vertical temperature gradient at intermediate heights in the disk 
atmosphere: a larger $\chi$ leads to a cooler temperature in a fixed vertical 
layer.  Although potentially important for more direct tracers of the disk 
atmosphere (e.g., scattered light or molecular line emission), these effects 
are indistinguishable from adjustments to the scale height parameters, \{$h_c$, 
$\psi$\}, with the data used here.

% TABLE 4
\begin{deluxetable*}{lcccc|cccc|cc|cc}
\tablecolumns{13}
\tablewidth{0pt}
\tabletypesize{\small}
\tablecaption{Model Parameters\label{structure_table}}
\tablehead{
\colhead{Name} & \colhead{$M_d$} & \colhead{$R_c$} & \colhead{$H_{100}$} & \colhead{$\psi$} & \colhead{$R_{\rm cav}$} & \colhead{$\log{\delta_{\rm cav}}$} & \colhead{$H_{\rm rim}$} & \colhead{$H_{\rm wall}$} & \colhead{$s_{\rm max}$} & \colhead{$p$} & \colhead{$i$} & \colhead{PA} \\
\colhead{} & \colhead{[M$_{\odot}$]} & \colhead{[AU]} & \colhead{[AU]} & \colhead{} & \colhead{[AU]} & \colhead{} & \colhead{[R$_{\ast}$]} & \colhead{[AU]} & \colhead{[$\mu$m]} & \colhead{} & \colhead{[\degr]} & \colhead{[\degr]} \\
\colhead{(1)} & \colhead{(2)} & \colhead{(3)} & \colhead{(4)} & \colhead{(5)} & \colhead{(6)} & \colhead{(7)} & \colhead{(8)} & \colhead{(9)} & \colhead{(10)} & \colhead{(11)} & \colhead{(12)} & \colhead{(13)}}
\startdata
MWC 758         & 0.008 &  25 & 18.3 & 0.05 & 73 & $-5.3$ &                      4.6 & 9.5 & 0.25 & 3.5 & 23 & 49  \\
SAO 206462      & 0.026 &  55 & 9.6  & 0.15 & 46 & $-5.2$ &                      1.6 & 9.2 & 3    & 3.5 & 12 & 64  \\
LkH$\alpha$ 330 & 0.024 &  60 & 6.5  & 0.20 & 68 & $-5.2$ &                      4.1 & 6.8 & 1    & 2.5 & 35 & 80  \\
SR 21           & 0.006 &  15 & 7.6  & 0.15 & 36 & $-5.9$ &                      0.8 & 8.2 & 10   & 3.5 & 22 & 100 \\
UX Tau          & 0.007 &  20 & 4.1  & 0.35 & 25 & $-5.7$ &                      0.5 & 0.6 & 1    & 2.5 & 35 & 176 \\
SR 24 S         & 0.045 &  40 & 4.5  & 0.20 & 29 & $-5.4$ &                      3.2 & 1.5 & 0.25 & 3.5 & 45 & 24  \\
DoAr 44         & 0.007 &  25 & 3.1  & 0.02 & 30 & $-5.1$ &                      3.1 & 9.0 & 0.5  & 3.5 & 35 & 80  \\
LkCa 15         & 0.055 &  85 & 2.9  & 0.20 & 50 & $-6.0$ &                      1.0 & 7.0 & 0.25 & 3.5 & 49 & 241 \\
RX J1615-3255   & 0.128 & 115 & 3.4  & 0.25 & 30 & $-5.8$ & \nodata\tablenotemark{a} & 2.0 & 0.25 & 3.5 & 41 & 143 \\
GM Aur          & 0.070 & 120 & 6.2  & 0.35 & 28 & $-5.7$ &                      0.1 & 2.8 & 0.5  & 3.5 & 55 & 64  \\
DM Tau          & 0.040 & 135 & 4.2  & 0.20 & 19 & $-4.8$ & \nodata\tablenotemark{a} & 5.7 & 3    & 3.5 & 35 & 155 \\
WSB 60          & 0.028 &  30 & 10.6 & 0.10 & 15 & $-1.7$ &                      0.4 & 0.8 & 1    & 3.5 & 28 & 117
\enddata
\tablecomments{Col.~(1): Name of host star.  Col.~(2): Total mass in the disk
(gas+dust, assuming a gas-to-dust mass ratio of 100:1).  Col.~(3):
Characteristic radius of the surface density profile.  Col.~(4): Scale height
at 100\,AU (based on $h_c$).  Col.~(5): Power-law index of the radial scale
height distribution.  Col.~(6): Radius of the dust-depleted disk cavity.
Col.~(7): Density contrast in the cavity, such that $\delta_{\rm cav} =
\Sigma/\Sigma_g$ when $R \le R_{\rm cav}$, where $\Sigma_g$ is the surface
density profile for a continuous disk (see Equation 1 and \S 3.1).  Col.~(8):
Scale height of the inner rim at the sublimation radius (expressed in units of
the stellar radius).  Col.~(9): Scale height of the cavity wall.  Col.~(10):
The maximum grain size for the dust population in the cavity and wall.
Col.~(11): The power-law index of the grain size distribution in the cavity and
wall, such that $n(s) \propto s^{-p}$.  Col.~(12): Disk inclination (0\degr\ is
face-on).  Col.~(11): Position angle of the disk major axis, measured east of
north.}
\tablenotetext{a}{We have removed any dust out to a radius of 1\,AU for the DM
Tau disk and 0.5\,AU for the RX J1615-3255 disk.  Neither case requires a
puffed-up inner rim to fit their infrared SED emission (see the Appendix).}
\end{deluxetable*}

\subsection{Modeling Synopsis}

The previous sections describe in detail an effort to model the SEDs and 
880\,$\mu$m continuum visibilities for this sample of transition disks in a 
homogeneous, parametric framework.  Before presenting those results, we first 
summarize some fundamental aspects of the modeling process:
\begin{itemize}
\item Using two-dimensional, axisymmetric Monte Carlo radiative transfer 
simulations, a parametric dust density structure is irradiated by a model star 
to determine an internally consistent temperature structure.  The results of 
the simulation are used to generate synthetic data products to compare with 
observations.  The underlying structure model is motivated by a simple viscous 
accretion disk model, with significant dust depletion in a central cavity.
\item Given the relatively faint infrared excesses for transition disks, the 
adopted stellar input spectrum factors significantly into inferences about 
material inside the cavity. 
\item Small quantities of (small) dust in the cavity can produce large amounts 
of infrared emission.  However, the lack of sufficient angular resolution in 
those regions means that the detailed structure of that material is unknown; 
considerable uncertainties remain for the mass content, spatial structure, and 
dust properties inside the cavity ($R \le R_{\rm cav}$).  Our estimates of 
density depletion tends to be maximal ($\delta_{\rm cav}$ is minimized): more 
material could reside in the inner disk if it has relatively lower infrared 
opacities (e.g., grain sizes larger than a few $\mu$m).  
\item Small, local modifications to the vertical distribution of material near 
the sublimation radius (the inner rim) and outer edge of the disk cavity (the 
cavity wall) are needed to explain the distinctive infrared 
spectra of the transition disks.  Because those same features can effectively 
shadow material at larger radii, their properties are at least partially 
entangled with one another as well as more global structure parameters.  
\item Because the dust-depleted cavities are spatially resolved at 880\,$\mu$m, 
their sizes are determined directly and with reasonable accuracy (typically 
within $\pm$10\% for a fixed $\gamma$).
\item Inherent model degeneracies and limited observational information make it 
difficult to quantify several model parameters for an individual disk with 
sufficient confidence.  However, because the modeling analysis was performed in 
a homogeneous framework for the entire sample, intra-sample (and qualitative) 
comparisons of the inferred structures are informative. 
\end{itemize}

% FIG 6
\begin{figure*}
\epsscale{1.1} 
\plotone{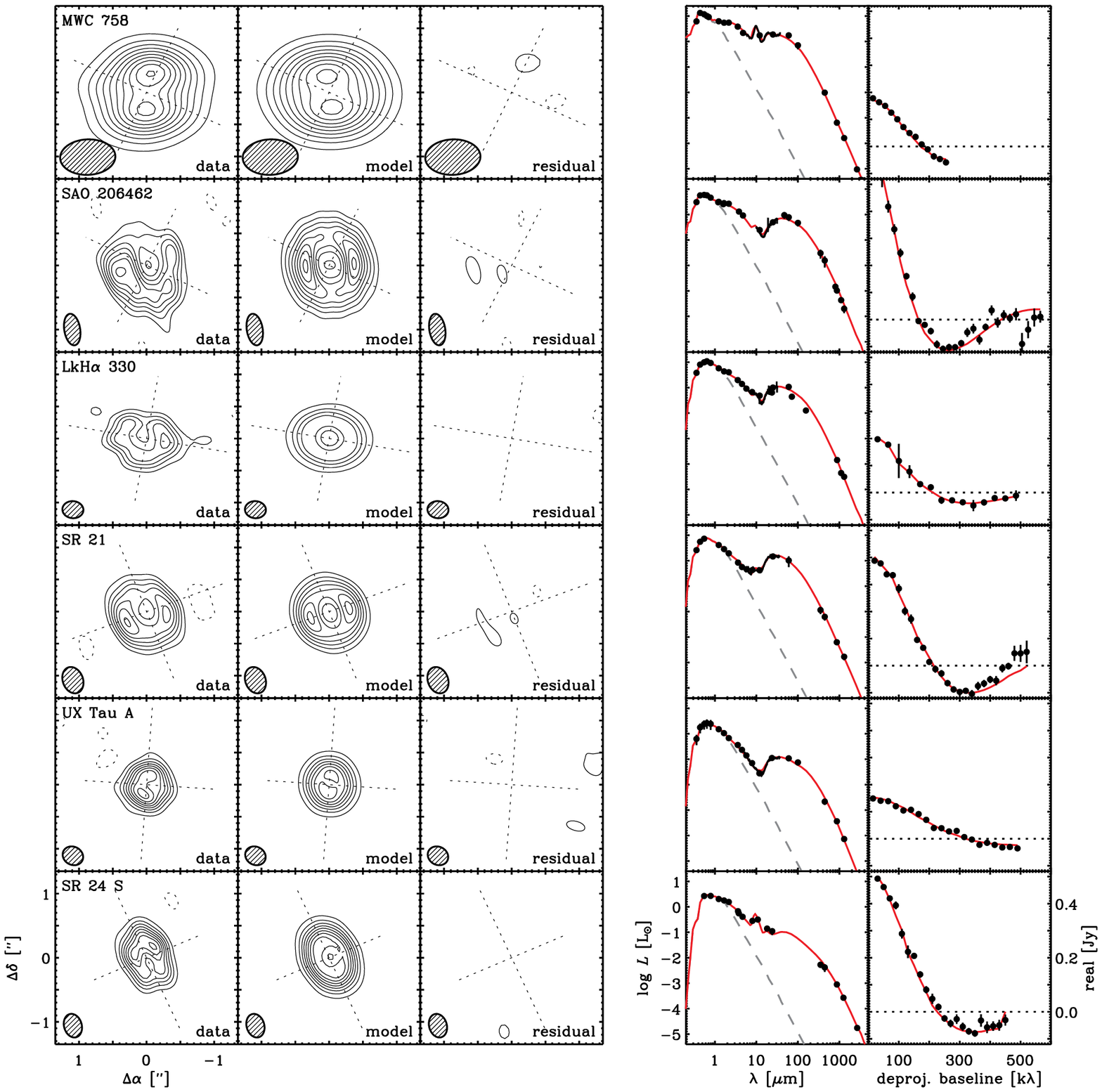} 
\figcaption{Comparisons of the disk structure models with observations.  The
left panels show the SMA continuum image, the model, and the imaged residuals,
with 3\,$\sigma$ contour intervals (see Table \ref{image_table}).  Crosshairs
mark the disk center (see \S 2 and Table \ref{obs_journal}) and major axis
orientation; their relative lengths represent the disk inclination.  The right
panels show the SEDs and 880\,$\mu$m visibility profiles, with models overlaid
in red.  The stellar photosphere models are marked with dashed curves.
\label{results1}}
\end{figure*}

% FIG 7
\begin{figure*}
\epsscale{1.1}
\plotone{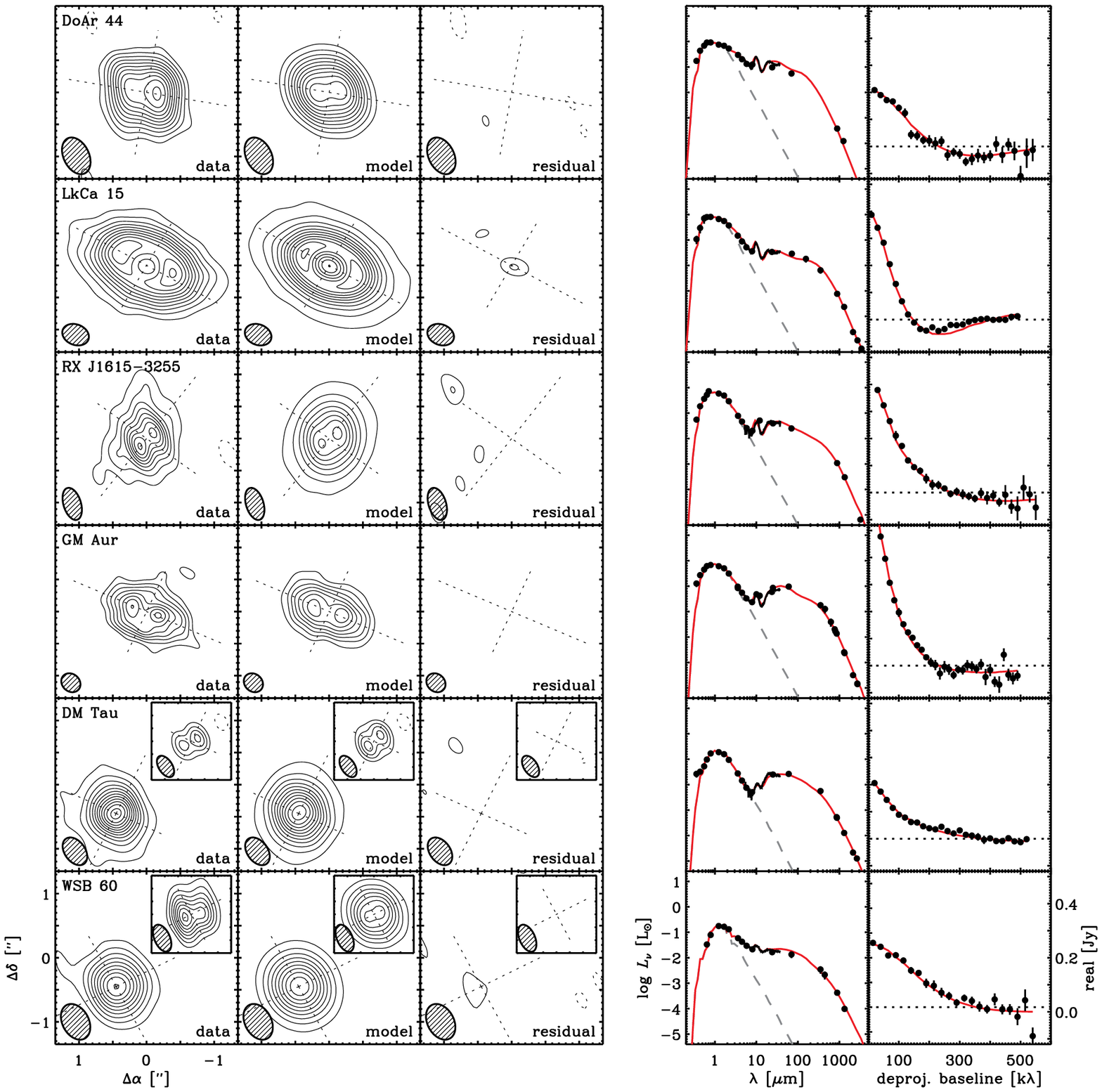}
\figcaption{Same as Figure \ref{results1}.  The images for DM Tau and WSB 60
are slightly offset to make room for the inset maps that were synthesized with
higher angular resolution (see Table \ref{image_table}).  \label{results2}}
\end{figure*}

\section{Results}

\begin{figure*}
\epsscale{1.0}
\plotone{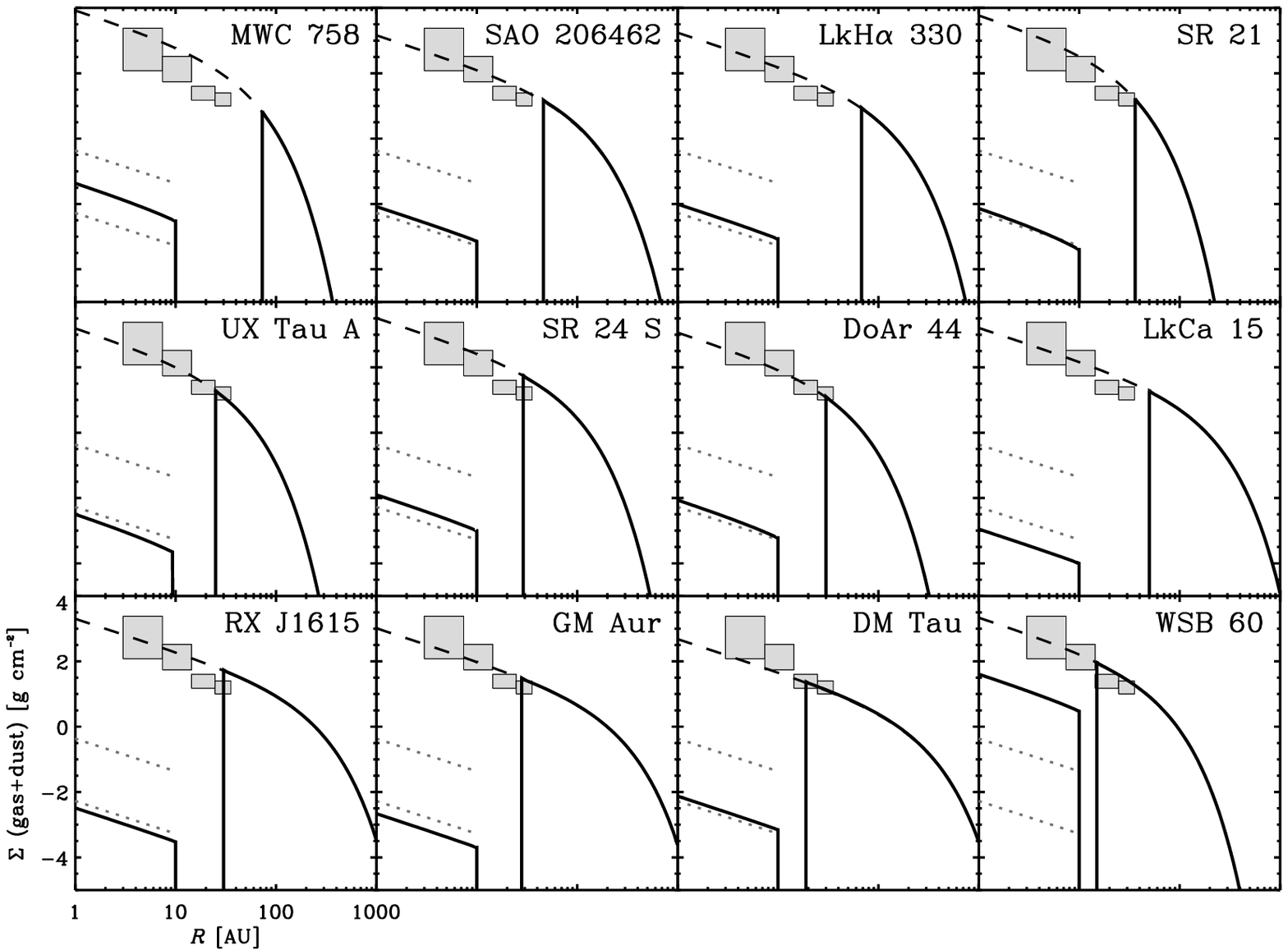}
\figcaption{Model surface density profiles (solid curves), assuming a
gas-to-dust mass ratio of 100:1.  The underlying global density profiles,
$\Sigma_g$ (see Eq.~1) are overlaid as dashed curves.  The $\Sigma$ gradient
parameter was fixed in the modeling to $\gamma = 1$, meaning that $\Sigma 
\propto 1/R$ at small radii ($R < R_c$) and $\Sigma \propto 1/e^R$ at large
radii ($R > R_c$).  Intended as references, the gray squares correspond to the
estimated surface densities in the Minimum Mass Solar Nebula (MMSN), a
representation of the primordial, solar-composition disk densities required to
account for the current giant planet masses in our solar system \citep[and
their uncertainties and feeding zones;][]{weidenschilling77}.  Dotted lines in
the inner disk mark the expected surface densities for 1 earth mass and 1 lunar
mass of material (gas + dust), assuming $\Sigma \propto 1/R$ from 0.1-10\,AU.
Note that the material inside the disk cavities is {\it not} resolved with our
SMA observations; the projected resolution limits correspond to $R \approx 
20$-40\,AU (depending on the distance to an individual system).  \label{sigma}}
\end{figure*}

\begin{figure*}
\epsscale{1.0}
\plotone{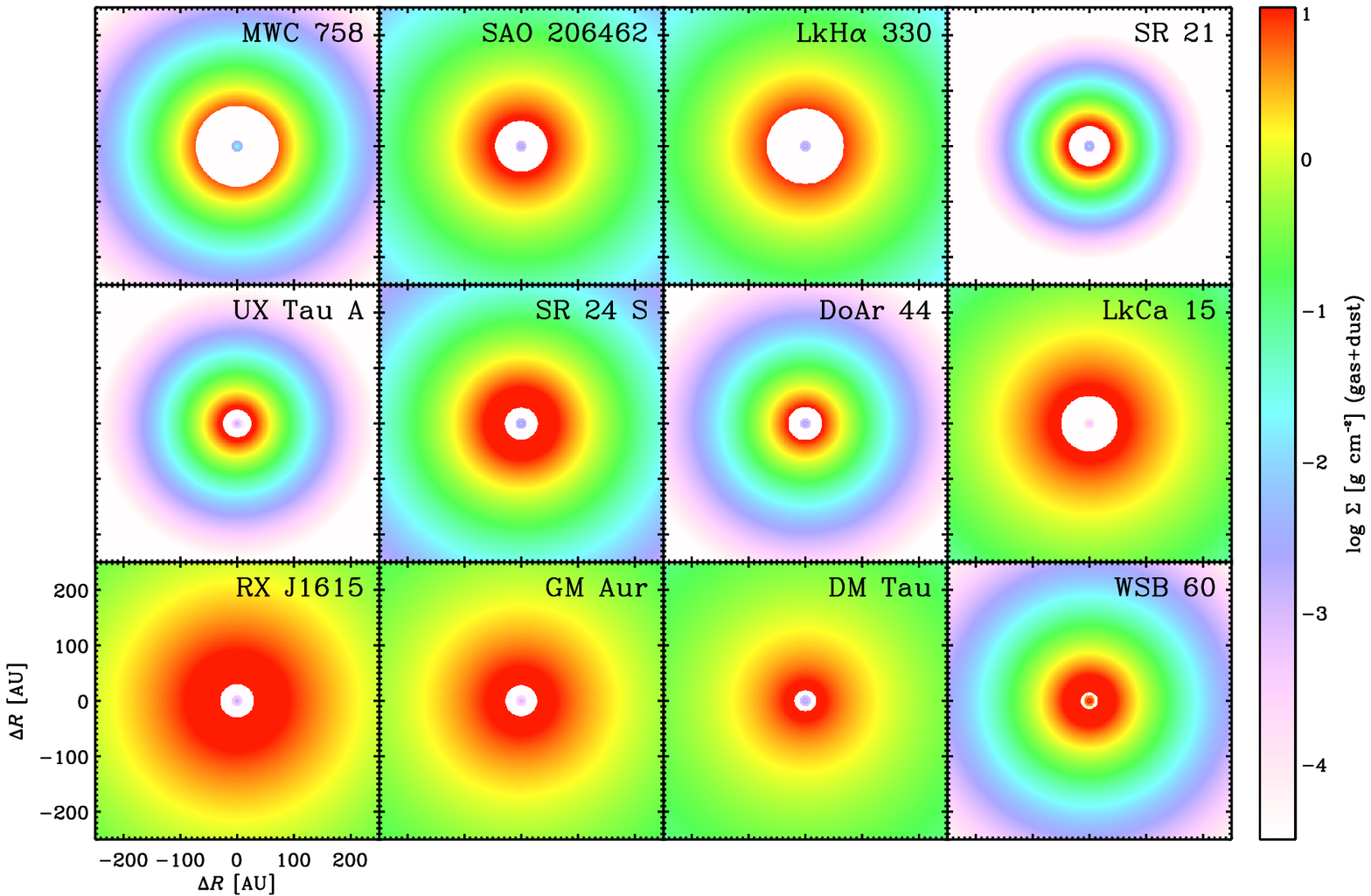}
\figcaption{The model surface densities in a map format are shown together with 
the same color stretch (a $\Sigma$ scale-bar is on the right).  This 
linear representation of the $\Sigma$ profiles provides a clear 
illustration of the range of cavity sizes, outer disk densities, and sizes 
($R_c$).  \label{spix}}
\end{figure*}

The estimated model parameters are compiled in Table \ref{structure_table}, and 
the data are compared with the modeling results in Figures \ref{results1} and 
\ref{results2}.  Each row in those figures corresponds to an individual disk, 
with the panels showing (from left to right) images of the data (as in Figure 
\ref{images}), model (synthesized in the same way as the data), and residuals 
(the imaged residual visibilities), the SED (data in black, model in red, 
stellar input spectrum as dashed curve) and the 880\,$\mu$m visibility profile 
(data in black, model in red).  In Table \ref{structure_table}, we have recast 
$\Sigma_c$ into a total disk mass, $M_d$, by integrating the surface density 
profile and assuming a gas-to-dust mass ratio of 100:1.  The scale height 
normalization ($h_c$) is expressed in physical units at a fixed radius of 
100\,AU for easy comparison.  The inner rim and cavity wall heights are also 
expressed in physical units, the former in terms of the fraction of the stellar 
radius to suggest its potential for shadowing material at 
larger radii.  The dust parameters correspond only to the material inside the 
cavity and wall (the dust populations in the outer disk were fixed; see \S 
3.2): for reference, their values in the ISM are $s_{\rm max} = 0.25$\,$\mu$m 
and $p = 3.5$ \citep{draine84,weingartner01}.  The corresponding model surface 
density profiles are shown together in Figure \ref{sigma}, along with the 
underlying global profiles, $\Sigma_g$ (dashed; see Eq.~1).  It is important 
to keep in mind that the surface density gradient was fixed, so all profiles 
have $\gamma = 1$ by default.  In practice, this means that the inner disk has 
$\Sigma \propto 1/R$, while the outer disk is dominated by the exponential 
taper, $\Sigma \propto 1/e^R$ (when $R_c > R_{\rm cav}$, part of the outer disk 
$\Sigma$ profile flattens out to vary inversely with radius; e.g., GM Aur).  
The surface densities are also shown in an alternative ``map" format in Figure 
\ref{spix}, which better highlights the range of cavity sizes on a linear 
radius scale. 

The transition disks in this sample have remarkably large dust-depleted 
cavities, with a range of radii ($R_{\rm cav}$) from 15-73\,AU.  The 
sample is likely biased toward large cavities, as they tend to produce a more 
obvious ``dip" in the infrared SED.  However, the extent of that bias is 
unclear, as some disks do not show any obvious signatures of a cavity in their 
infrared spectra.  The WSB 60 and MWC 758 disks are good examples; their 
cavities were discovered serendipitously in millimeter-wave images 
\citep{andrews09,isella10b}.  Efforts to infer $R_{\rm cav}$ from the 
unresolved SED alone can be extraordinarily challenging (see \S 3.5).  However, 
there have been some notable successes when compared with direct imaging, 
including GM Aur \citep{calvet05,hughes09}, SAO 206462 \citep[][and this 
paper]{brown07} and the older disk around TW Hya \citep{calvet02,hughes07}.  
While sophisticated SED modeling does surprisingly well in many cases $-$ 
typically agreeing with the $R_{\rm cav}$ values determined here within 
$\sim$30\%\ \citep[e.g.,][]{brown07,espaillat07} $-$ there are also some cases 
where the predicted cavity sizes are much less consistent with the SMA images.  
The DM Tau disk is the most extreme example, where the 3\,AU cavity inferred by 
\citet{calvet05} is $\sim$6$\times$ smaller than measured here ($R_{\rm cav} = 
19$\,AU).  Such large discrepancies are informative; they call for 
modifications to the assumed structure and/or dust properties inside the 
cavity.  For example, \citet{espaillat10b} were able to decrease their original 
estimate of the cavity size in the UX Tau A disk by more than a factor of 2 
\citep[from 71 to 30\,AU;][]{espaillat10} using an updated grain composition in 
the cavity wall, both improving their SED-based fit quality and providing 
better agreement with the resolved SMA data.  These more extreme examples 
should serve as reminders that estimates of cavity sizes from SEDs are 
model-dependent; they should not be considered robust until the cavity is 
spatially resolved.  

% FIG 8

% FIG 9

The apparent emptiness of the transition disk cavities is a striking aspect of 
the modeling results.  Compared to the underlying global $\Sigma_g$ profiles 
(dashed curves in Figure \ref{sigma}), the surface densities inside these 
cavities have been decreased by $\sim$5-6 orders of magnitude.  The WSB 60 disk 
is a notable exception, where a more substantial amount of dust is required to 
account for the infrared excess around this cool star (the $\Sigma$ decrease is 
a factor of $\sim$50, $\delta_{\rm cav} = 0.02$).  In most cases, only a few 
percent of a {\it lunar} mass of dust ($\sim$10$^{-10}$-10$^{-8}$\,M$_{\odot}$) 
is required to reproduce the observed infrared excesses.  As explained in \S 
3.5, the dust depletion factors ($\delta_{\rm cav}$; and therefore $\Sigma$ in 
the inner disk) were effectively determined from the infrared spectrum, and 
therefore are only relevant for small ($\sim$$\mu$m-sized) dust grains in the 
cavities.  Increased densities (larger $\delta_{\rm cav}$) could easily be 
accomodated for a population of larger dust grains that have lower infrared 
opacities (i.e., $s_{\rm max} \gg 1$\,$\mu$m).  We can crudely estimate the 
mass of mm-sized grains inside the cavities from the 880\,$\mu$m residual maps 
in Figures \ref{results1} and \ref{results2}.  Since our models produce no 
millimeter emission inside $R_{\rm cav}$, the residual emission near the disk 
center should roughly trace the product of the dust opacity, inner disk mass, 
and a mean temperature \citep[$F_{\nu} \propto \kappa M_{\rm cav} \langle T 
\rangle$; e.g.,][]{aw05}.  In most cases, those residuals are below the noise 
floor.  Using 3\,$\sigma$ upper limits from the residual map centers (see Table 
\ref{image_table}), an 880\,$\mu$m opacity of 3.6\,cm$^2$ g$^{-1}$ (see \S 
3.2), and a mean temperature of 75\,K, the absence of significant residuals 
suggests that there is typically less than $\sim$1\,M$_{\oplus}$ ($\lesssim 
10^{-6}$\,M$_{\odot}$) of mm-sized grains in these disk cavities.\footnote{Note 
that the constraint from the 880\,$\mu$m residuals in this simple example is 
technically less stringent than could be made from the infrared emission.  The 
1\,$\mu$m opacities for the large grain population, \{$s_{\rm max} = 1$\,mm, $p 
= 3.5$\}, are only $\sim$50$\times$ lower than for $s_{\rm max} = 1$\,$\mu$m 
(see Figure \ref{opacities}), meaning these ``mm-sized" grains would produce 
too much infrared radiation {\it before} they create 880\,$\mu$m residuals 
(when $M_{\rm cav} \lesssim 10^{-8}$\,M$_{\odot}$).  Nevertheless, the example 
is instructive in pointing out that the 880\,$\mu$m data are not quite 
sensitive enough to provide very strong limits on $\delta_{\rm cav}$.}  Due to 
our sensitivity limits, these 880\,$\mu$m constraints on the density contrasts 
are relatively weak ($\delta_{\rm cav} \le 0.01$-0.1).  The LkCa 15 disk is a 
remarkable exception, as it shows a significant ($\sim$6\,$\sigma$; $F_{\nu} 
\approx 10$\,mJy) and centralized residual peak that suggests a small reservoir 
($\sim$1\,M$_{\oplus}$) of mm-sized grains are present inside its cavity. 

Aside from the large dust-depleted cavities, these transition disks have 
surface density profiles similar to those inferred for normal disks 
\citep[see][]{andrews09,andrews10b}.  Both populations exhibit the same range 
of characteristic scaling radii and disk masses, with $R_c = 15$-135\,AU and 
$M_d = 0.006$-0.128\,M$_{\odot}$ inferred here (assuming a 100:1 gas-to-dust 
mass ratio).  Surprisingly, the transition disks are distributed along the same 
$M_d$-$R_c$ correlation that was noted for normal disks in Ophiuchus 
\citep{andrews10b}, with essentially the same scatter.  Of course, disk mass 
comparisons between these two populations are misleading, as the cavities in 
transition disks are substantial mass reservoirs in normal disks.  The same 
disk models {\it without} dust-depleted cavities (i.e., if $\Sigma = \Sigma_g$ 
at all radii) would be $\sim$1.3-3$\times$ more massive than indicated in Table 
\ref{structure_table} ($\sim$20$\times$ for the disk with the largest cavity, 
MWC 758).  This implies that the surface densities in the outer disk are 
typically up to a few times higher in transition disks 
compared to their normal disk counterparts.  

With a relatively small and potentially heavily-biased sample, there is little 
that can be concluded from any trends between the model structure parameters.  
As noted above, the outer parts of the transition disks do seem to follow the 
tentative $M_d$-$R_c$ correlation identified from a sample of normal T Tauri 
disks \citep{andrews10b}.  With a brief examination of the figures here (with 
the disks ordered by spectral classification), one might notice a possible 
trend between the stellar type and cavity size.  Indeed, there is a marginal 
hint of a positive correlation ($\sim$2.5\,$\sigma$) between $R_{\rm cav}$ and 
$T_{\rm eff}$ (also $L_{\ast}$ and $M_{\ast}$), such that the earlier type 
stars appear to have disks with larger cavities.  However, the trend is likely 
an artifact of the endpoints being outliers: the low-mass stars host disks with 
small cavities (DM Tau, WSB 60), and some of the higher-mass stars have disks 
with larger ones (MWC 758, LkH$\alpha$ 330), but there is a wide range of 
$R_{\rm cav}$ for intermediate cases.  No other trends among the disk model 
parameters, or between them and the stellar properties, are found here.  A more 
individualized description of the model results for each disk, along with a 
brief comparison to previous models when available, can be found in the 
Appendix.

\section{Discussion}

\subsection{Transition Disks and Evolution}

\subsubsection{Nomenclature and Our Definition of Transition Disk}

Before integrating this sample of transition disks into a larger evolutionary 
context, it seems necessary to comment on its placement in the rather 
disorienting landscape of ``evolved" disks in the current literature.  Some of 
that confusion is related to nomenclature, and some is due to the inherently 
subjective nature of empirical classification.  Our definition of a 
``transition" disk is conservative: we mean only those cases with 
unequivocal evidence for a sharp drop in the dust optical depths inside some 
radius, either in the form of a dip and then dramatic rise in the infrared 
spectrum over a narrow wavelength interval and/or a clearly resolved cavity at 
the disk center.  Various other types of disks have been associated with the 
transition phase, including cases without a distinctive dip in the infrared.  
Such disks - sometimes termed ``anemic" or ``homologously-depleted" - have 
relatively faint infrared excesses and often low accretion rates, but do not 
(yet) exhibit any compelling observational evidence for a dust-depleted cavity 
\citep{lada06,najita07,cieza08,cieza10,currie09a,currie09b}.  While they may 
signal an interesting evolution phase in their own right, these anemic disks 
are not {\it bona fide} transition disks in the sense intended here.  Instead, 
the transition disks (in our adopted meaning) are typically identified with a 
set of specialized infrared color criteria 
\citep{cieza07,cieza10,fang09,muzerolle10,merin10}.  However, none of these 
color-selection methods successfully recovers our entire sample (with failure 
rates up to 60\%), despite the unambiguous confirmations of large disk cavities 
from the SMA (the Mer{\'{\i}}n et al.~``cold disk" criteria perform well; a 
slight expansion of their selection-space would encompass nearly all the disks 
in this sample).  

These low recovery rates are not surprising, especially given the sometimes 
bright (and complex) emission spectrum generated by a small amount of dust 
inside the disk cavities.  Due to the bewildering diversity of possible 
infrared SED behaviors for transition disks, any (unresolved) infrared-based 
identification scheme is biased and will miss good candidates.  Unfortunately, 
our current ignorance of the material and structural properties of the inner 
disk makes it difficult to properly characterize those biases.  In the end, the 
only appropriate systematic way to find and characterize the transition disk 
population is with a flux- and resolution-limited survey that can directly 
image dust-depleted cavities.  Since that type of unbiased survey is not yet 
feasible, we focus on the small piece of selection-space that is (at least 
partially) addressed by the specific type of transition disks in this sample: 
namely, those with {\it bright} millimeter emission and {\it large} cavities.

\subsubsection{Frequency of Transition Disks with Large Cavities}

\begin{figure}
\epsscale{1.1}
\plotone{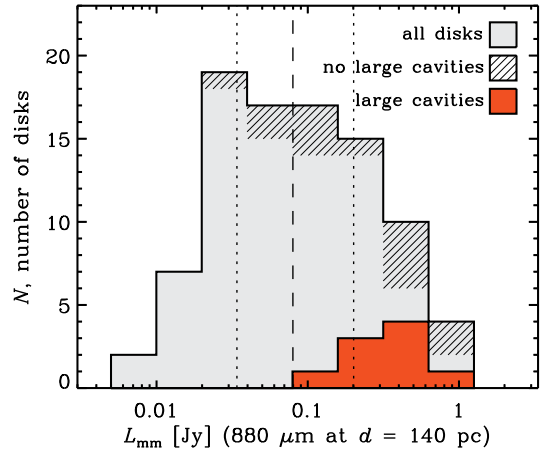}
\figcaption{The distribution of 880\,$\mu$m luminosities ($L_{\rm mm}$), scaled 
to a distance of 140\,pc, for disks in the Taurus and Ophiuchus star-forming 
regions (grey), excluding non-detections, sources with potential envelope 
contanimation (including flat-spectrum objects), and early-type (A/B) stars 
\citep[data from][]{aw05,aw07b}.  The median and quartiles of the distribution 
are shown as dashed and dotted lines, respectively.  The $L_{\rm mm}$ 
distribution of the subsample of Tau/Oph millimeter-bright transition disks 
with directly resolved large ($R_{\rm cav} \gtrsim 15$\,AU) cavities is shown 
in red.  The hatched regions correspond to disks that do not have large 
cavities, as determined from millimeter-wave observations with comparable 
resolution to this study \citep{andrews09,andrews10b}.  In each bin, the 
fraction of transition disks with large cavities is at least $N$(red)/$N$(grey) 
and at most $N$(hatched)/$N$(grey).  That fraction is at least 20\%\ (and at 
most 70\%) for the upper half of the $L_{\rm mm}$ distribution, and at least 
26\%\ (at most 70\%) for the upper quartile.  As indicated by the large number 
of disks that lie in the unhatched grey areas of the $L_{\rm mm}$ distribution, 
many disks have not yet been observed at sufficient resolution to detect a 
dust-depleted cavity.  Moreover, we do not yet know the transition disk 
fraction at low $L_{\rm mm}$ (below the median) or smaller $R_{\rm cav}$.   
\label{Lmm_dist}}
\end{figure}

Despite the narrowed scope, transition disks with large cavities are 
surprisingly common when viewed in the proper context.  
Consider the millimeter-wave luminosities from disks in the 
nearby Tau and Oph associations \citep{aw05,aw07,cieza10}.  Removing 
non-detections, sources with likely envelope contamination (including 
flat-spectrum objects), and disks around early-type (A/B) stars, and scaling 
1.3\,mm photometry when necessary by a conservative factor $1/\lambda^3$ 
(steeper than the typical slope found by Andrews \& Williams), we constructed 
the distribution of 880\,$\mu$m luminosities ($L_{\rm mm}$) shown in Figure 
\ref{Lmm_dist} for 91 disks.  Out of that sample, large cavities ($R_{\rm cav} 
\gtrsim 15$\,AU) have been directly resolved for 9 disks \citep[10\%, including 
RY Tau; see][]{isella10}.  That fraction is roughly consistent with 
expectations from infrared-selection methods \citep{muzerolle10}, especially if 
the upper limits for millimeter-faint disks are included.  But current disk 
imaging surveys are sensitivity-limited and have only probed the brightest 
disks.  In the upper half of the $L_{\rm mm}$ distribution ($\ge 80$\,mJy at 
140\,pc), the transition disks with large cavities comprise a larger fraction, 
$\sim$20\%\ (9/46 disks).  And in the upper quartile of the distribution ($\ge 
200$\,mJy at 140\,pc), that fraction is slightly higher still (26\%; 6/23 
disks).  These transition disk fractions are lower bounds; there could very 
well be more disks with large cavities that have yet to be imaged at sufficient 
resolution.  The maximum fraction of this type of transition disk in the upper 
half (or quartile) of the $L_{\rm mm}$ distribution is $\sim$70\%, based on the 
known absence of large cavities in some of the remaining disks 
\citep{andrews09,andrews10b}.  

% FIG 10

To reiterate those remarkable statistics: at least 1 in 5, and possibly up 
to $\sim$two-thirds of the known millimeter-bright disks have {\it large} 
dust-depleted cavities (with $R_{\rm cav} \gtrsim 15$\,AU).  We do not know how 
high such fractions might be for smaller cavities, or disks with even fainter 
millimeter emission, or young disk populations in other environments (if these 
results for Tau/Oph are to be applied to other clusters, one must keep in mind 
the counting uncertainties that correspond to $\pm$10\%\ errors on the 
transition disk fractions quoted above).  Regardless, the implication is that a 
large fraction of the most millimeter-bright -- and therefore most massive -- 
disks show unequivocal evidence for substantial dust evolution in the regions 
most relevant for planet formation on timescales of less than $\sim$1-2\,Myr 
(for reference, the median and quartile $L_{\rm mm}$ values correspond to disk 
masses of 0.005\,M$_{\odot}$ and 0.011\,M$_{\odot}$, respectively, using the 
standard Andrews \& Williams assumptions).  This reinforces a similar 
conjecture made by \citet{najita07} based on a group of unresolved Taurus 
disks, although see \S 5.3.3 regarding their corollary claim of low accretion 
rates.

\subsubsection{Comments on Evolution}

In retrospect, it is striking that these transition disks could be such bright 
millimeter sources in spite of their depleted cavities.  What were these disks 
like {\it before} they evolved, and are such progenitor disks still present in 
these young associations?  If we assume that these disks had $\Sigma = 
\Sigma_g$ at all radii before depleting their cavities, our radiative transfer 
calculations indicate that $L_{\rm mm}$ would typically be $\le$2$\times$ 
larger (up to 4$\times$ for the extreme case, MWC 758).  Note that the emission 
increase is not directly proportional to the mass increase expected from 
filling the cavity with dust (see \S 4), due to optical depth effects.  
Unfortunately, there are only a few such disks on the extreme tail of the 
$L_{\rm mm}$ distribution.  With such low numbers, identifying a parent 
population for these massive transition disks is difficult.  It is possible 
that they represent the evolved versions of massive flat-spectrum sources 
(e.g., DG Tau, HL Tau).  Or perhaps some of them brightened {\it after} 
depleting their cavities: for example, viscous spreading of an initially narrow 
optically thick ring could increase $L_{\rm mm}$ \citep[if the flux scales with 
the disk size, $R_c$; see][]{andrews10b}.    

The cavity sizes and millimeter luminosities are not the only clues available 
to help characterize transition disk evolution, and we would be remiss not to 
address potential hints from the cavity contents.  In terms of their 
near-infrared excess luminosities, it is clear that not all transition disks 
are alike: some are faint (in this sample, DM Tau, GM Aur, RX J1615-3255, and 
perhaps SR 21) and others are quite bright (e.g., DoAr 44).  
\citet{espaillat07} suggested that the latter cases, dubbed ``pre-transition" 
disks, may be more representative of gaps than cavities.  That hypothesis (and 
accompanying nomenclature) supposes that the transition disks with brighter 
infrared excesses trace an earlier stage of the inner disk depletion process.  
That may very well be the case, but the properties of the inner disk are so 
poorly constrained that it is not the only possibility.  While there is no 
clear association of infrared excess luminosity with the depletion factor, 
$\delta_{\rm cav}$, we were forced to remove dust inside $\sim$1\,AU to avoid 
over-producing the observed emission for our DM Tau and RX J1615-3255 models 
(see the Appendix).  That is in good agreement with the original 
\citet{espaillat07} conjecture, where the inner disk densities (or optical 
depths) and excess levels are related.  However, the strength of the infrared 
excess is also tied to the vertical distribution of dust near the star: larger 
heights generate more emission \citep[see \S 3, or][]{espaillat10}.  Perhaps 
the transition disks with 
brighter infrared emission also have more ``puffed"-up inner rims.  This might 
suggest that those cases with faint excesses have relatively lower inner disk 
{\it gas} densities (or gas-to-dust ratios), which do not provide sufficient 
pressure to support such extended vertical structures.

\subsection{Potential Dust Depletion Mechanisms}

Up to this point, the discussion has been intentionally generic with respect to 
the physical origins of the dust-depleted cavities in transition disks.  In a 
broad sense, there are two types of depletion mechanisms: those that physically 
remove or ``clear" material from the inner disk (decrease $\Sigma$), and those 
that modify the emission properties of the dust (decrease $\kappa$).  Various 
inner disk evolutionary processes have been assessed in previous studies of 
transition disks \citep{dalessio05,najita07,alexander09}.  We review 
several of the proposed dust depletion mechanisms in the context of the 
specific type of transition disk in our sample.

\subsubsection{Dispersal by Winds}

Among the proposed clearing mechanisms, mass-loss in a photoevaporative wind 
has excellent potential for clearing a disk cavity 
\citep{clarke01,alexander06a,alexander06b}.  Energetic photons from the central 
star can ionize the disk surface layers and launch a wind from just inside the 
radius where the thermal energy of the ionized gas exceeds the escape velocity 
\citep[$R \approx 1$-10\,AU;][]{shu93,hollenbach94}.  If the inward mass 
(accretion) flow rate exceeds the mass-loss rate in this wind, changes to the 
disk structure occur slowly, as expected for viscous evolution.  But as the 
accretion flow decays over time, it will eventually be surpassed by the 
mass-loss rate in the wind \citep[assuming the photoionizing flux from the star 
remains high;][]{alexander04,alexander05}.  At that point, the accretion flow 
is redirected into the wind and gas is rapidly depleted in a gap near the wind 
launch radius.  Isolated from its replenishment source (the outer disk), the 
inner disk material quickly drains onto the star.  The initially small cavity 
made in this way exposes the disk edge near the wind launch radius to 
unattenuated high-energy irradiation, accelerating the dispersal of material 
and growth of the cavity.  Although this photoevaporation process only directly 
affects the gas phase of the disk (only the very smallest dust grains, $s \le 
0.1$\,$\mu$m, are directly swept away in the wind), \citet{alexander07} 
demonstrated that a simple prescription for the radial drift of solids promotes 
the simultaneous depletion of dust inside the wind-blown cavity.

In models that consider only UV irradiation, the wind plays a minor role in the 
evolution of a disk structure over most of its lifetime 
\citep{alexander06b,gorti09b}.  Only late in life, when the disk mass and 
accretion flow have diminished, does photoevaporation dramatically influence 
the disk properties.  Observationally, the transition disks produced by UV 
photoevaporation should have empty cavities, low accretion rates ($\dot{M} \le 
10^{-10}$\,M$_{\odot}$ yr$^{-1}$) and low masses ($M_d \le 
0.005$\,M$_{\odot}$), and orbit relatively ``old" stars \citep[$\sim$few 
Myr;][]{alexander07,alexander09}.  Given the large disk masses (Table 
\ref{structure_table}) and high accretion rates (Table \ref{stars_table}) in 
this sample, it is clear that UV photoevaporation is not responsible for 
creating the large cavities in these millimeter-bright transition disks 
\citep[as noted by][]{alexander07}.  However, there is growing consensus that 
X-rays drive substantially higher wind mass-loss rates, and therefore that 
photoevaporation has a much more prominent effect on disk structures at early 
times \citep{ercolano08,ercolano09,drake09,owen10a}.  The early onset of X-ray 
photoevaporation implies that these disks could still have material in their 
cavities, and may have higher accretion rates (up to $10^{-8}$\,M$_{\odot}$ 
yr$^{-1}$) and disk masses \citep[$M_d \sim 0.01$\,M$_{\odot}$;][]{owen10b}.  
To assess the role of X-ray photoevaporation, we compare the quiescent X-ray 
luminosities ($L_X$ in Table \ref{stars_table}) with the cavity sizes ($R_{\rm 
cav}$), disk masses, and accretion rates.  In this sample, the $R_{\rm cav}$ 
and $\dot{M}$ values for a given $L_X$ are much too large to be consistent with 
their formation via photoevaporative mass-loss \citep[as noted by][]{owen10b}.  
That is not to say that photoevaporation is unable to clear disk cavities, as 
it seems a likely explanation for {\it some} evolved disks 
\citep[e.g.,][]{cieza08,pascucci09} - only that it is insufficient to explain 
the population of massive transition disks with large cavities.  

Some alternative wind sources have been suggested as potential contributors to
the clearing of disk material.  Based on their local MHD simulations of
turbulence driven by the magneto-rotational instability (MRI), \citet{suzuki09} 
postulated that the disruption of large-scale channel flows by magnetic 
reconnection events can drive a substantial wind off the disk surface.  
Extrapolated to global evolution models for the disk structure, the mass-loss 
rates in the inner disk from such a wind could rival those produced by X-ray 
photoevaporation \citep{suzuki10}.  Those initial calculations suggest that 
$R_{\rm cav} \lesssim 15$\,AU cavities could be depleted on Myr timescales, but 
a more comprehensive set of observational predictions that can be used to test 
the feasibility of MRI disk winds as a transition disk clearing mechanism (and 
differentiate it from photoevaporation) is not yet available.  Aside from winds 
launched from the disk itself, the wind generated by the central star may 
similarly facilitate disk dispersal by stripping material out of the disk 
atmosphere \citep[e.g.,][]{horedt78}.  Recent models by \citet{matsuyama09} 
indicate that the mass-loss rate of disk material from this process is quite 
small for young disks, although perhaps the stellar wind can act in concert 
with disk winds to accelerate the dispersal of the inner disk.  That said, the 
potential synergy of these various wind dissipation models has not yet been 
explored in detail.

\subsubsection{Material Evolution - Particle Growth}

Instead of a lack of mass in the inner disk, the cavities we observe could 
instead be a sign that substantial {\it material} evolution has occurred in the 
inner parts of transition disks.  Solid particle growth is a natural disk 
process that might be able to explain the transition disk structures without 
requiring drastic density modifications.  Because large solid particles have 
low opacities ($\kappa$), significant grain growth can diminish the continuum 
emission over a wide wavelength interval \citep{miyake93,dalessio06}.  To 
produce the observed infrared SED deficits with a continuous density profile 
($\Sigma = \Sigma_g$), the $\sim$1-10\,$\mu$m opacities in the inner disk need 
to be decreased by a factor of roughly $\delta_{\rm cav}$ (because the infrared 
emission sets the contrast parameter; see \S 3.5).  The 880\,$\mu$m opacity 
also needs to decrease by a factor of $\ge$10-100, to account for the ring-like 
emission observed with the SMA.  With the dust composition assumed here, such 
opacity spectra are only possible after growth beyond $s_{\rm max} = 1$\,m 
(preferably with a top-heavy size distribution, $p \le 2.5$).  Coagulation 
timescales are expected to be shorter in the inner disk, due to the increased 
densities and relative velocities closer to the central star.  In principle, 
that would produce a radial variation in particle sizes and a decrease in 
$\kappa$ - and therefore emission - toward smaller disk radii.  

Nevertheless, there are challenges to reproducing the transition disk cavities 
with particle growth alone.  For example, models of the grain size evolution in 
disks predict smooth $\kappa$ variations 
\citep{dullemond05,garaud07,brauer08,birnstiel10}, rather than the {\it abrupt} 
optical depth change near $R_{\rm cav}$ that is observed.  To account for the 
sharp optical depth contrast signaled by the 880\,$\mu$m emission rings (i.e., 
the visibility nulls) and the shape of the mid-infrared spectrum, particle 
sizes would need to increase dramatically over a narrow radial range (at most a 
few AU).  That type of break in $\kappa(R)$ might be reproduced if particle 
growth is significantly enhanced in a large, quiescent dead zone with a sharp 
outer edge.  In a dense disk, the midplane can be so shielded from ionizing 
radiation that the MRI will not operate \citep{gammie96,sano00,fromang02}: 
without that source of disruptive turbulence, particle growth can proceed more 
efficiently inside the dead zone.  But there are two problems with this 
scenario.  First, the disk atmosphere will remain partially ionized and can 
channel material from the outer disk toward the star in surface accretion flows
\citep[e.g.,][]{gammie96}.  Even small amounts of $\mu$m-sized dust grains 
carried along in that flow could produce enough infrared radiation to wash out 
the distinctive dip in a transition disk SED.  And second, the cavities are 
much larger than is expected for dead zones \citep{matsumura03}.  Even in a 
large dead zone without dust in the surface layers, a collisional fragmentation 
process might still produce a substantial reservoir of small grains and limit 
growth beyond $\sim$cm-scales \citep{brauer08,birnstiel10,birnstiel11}.  New 
models of particle growth under dead zone conditions (with radiative transfer) 
are needed to determine if coagulation and fragmentation in a low-turbulence 
environment can produce an infrared spectrum and 880\,$\mu$m emission ring 
consistent with those observed for transition disks.  

Molecular emission tracers can provide additional insight here, as particle 
growth itself is not expected to significantly perturb the gas (H$_2$) 
densities inside the cavity.  The measured accretion rates (Table 
\ref{stars_table}; aside from SR 21) unambiguously confirm that neutral gas is 
flowing through these cavities and onto the stars.  Also, some of these 
transition disks emit CO rovibrational spectra that attest to the presence of 
warm molecular gas in their inner few AU 
\citep{najita03,salyk07,salyk09,pontoppidan08}.  However, new studies indicate 
that those inner regions are deficient in other molecules \citep[e.g., H$_2$O; 
see][]{najita10,pontoppidan10}, and \citet{dutrey08} claim that the GM Aur disk 
has a $R \approx 20$\,AU region of depleted CO emission.  In some respects, it 
is clear that transition disks have different molecular signatures compared to 
their normal disk counterparts.  But given the inherent challenges in 
interpreting such measurements, it remains uncertain whether their cavities 
also have depleted gas densities \citep[other effects like photodissociation, 
self-shielding, and gas-grain chemistry must also be considered; 
e.g.,][]{aikawa06,jonkheid07,vasyunin11}.  Nevertheless, we can expect that 
these molecular emission lines will be crucial in evaluating what role particle 
growth has in producing the observed dust structures of transition disks.

\subsubsection{Tidal Interactions with Companions}

A large fraction of stars have companions \citep{abt76,duquennoy91}, with 
stellar multiplicity fractions approaching 75\%\ in the diffuse young clusters 
near the Sun \citep[e.g.,][]{white01,kraus11}.  Circumstellar material is 
expected to be strongly affected by dynamical interactions with these stellar 
components \citep[see][]{lin93}.  The interplay of resonant and viscous torques 
can open a gap in a circum-system disk, truncate individual disk edges, and 
potentially expedite disk dissipation (via accretion or ejection), depending on 
the orbital separation ($a$), eccentricity ($e$), and mass ratio ($q$) of the 
companion.  Based on the calculations of \citet{artymowicz94}, a binary with a 
large separation can retain individual disks similar to those around single 
stars, but with their outer edges truncated at $R_o \sim 0.2$-0.5$a$.  For 
smaller separations, a circumbinary ring structure may also be present, with a 
truncated inner edge at $R_{cb} \sim 1.8$-2.6$a$.  \citet{pichardo08} find 
similar results, but generalize to arbitrary eccentricities such that $R_{cb} 
\sim 1.9 (1+e^{0.3})a$.  Their calculations suggest that the inner radius of 
the circumbinary ring has only a weak dependence on the mass ratio, at least 
for $q \ge 0.1$.  It is natural to associate the 880\,$\mu$m emission rings 
observed for the transition disks with these circumbinary structures.  In fact, 
stellar companions have been detected inside the cavities inferred (indirectly) 
for a few notable transition disks, including CoKu Tau/4 \citep{ireland08} and 
the older HD 98800 and Hen 3-600 systems \citep{uchida04,furlan07,andrews10a}.

% FIG 11
\begin{figure*}
\epsscale{1.1}
\plottwo{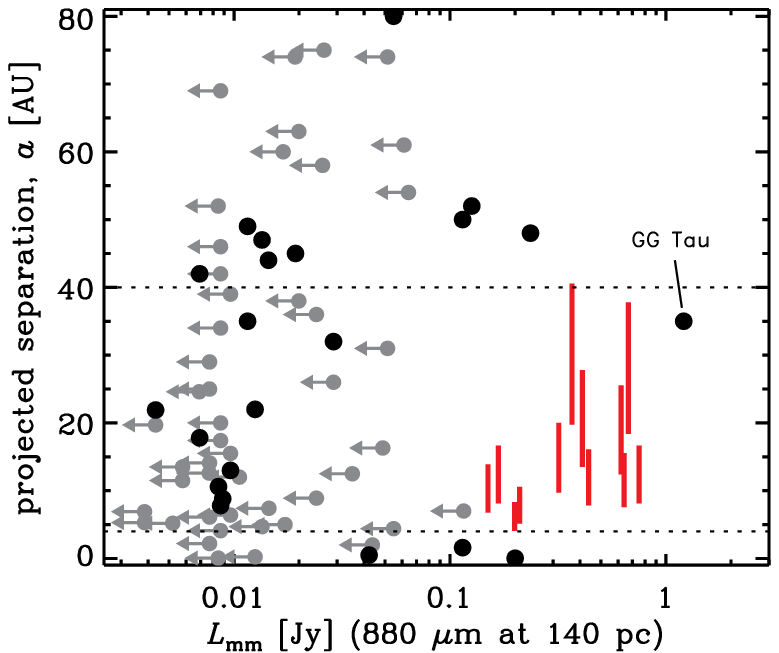}{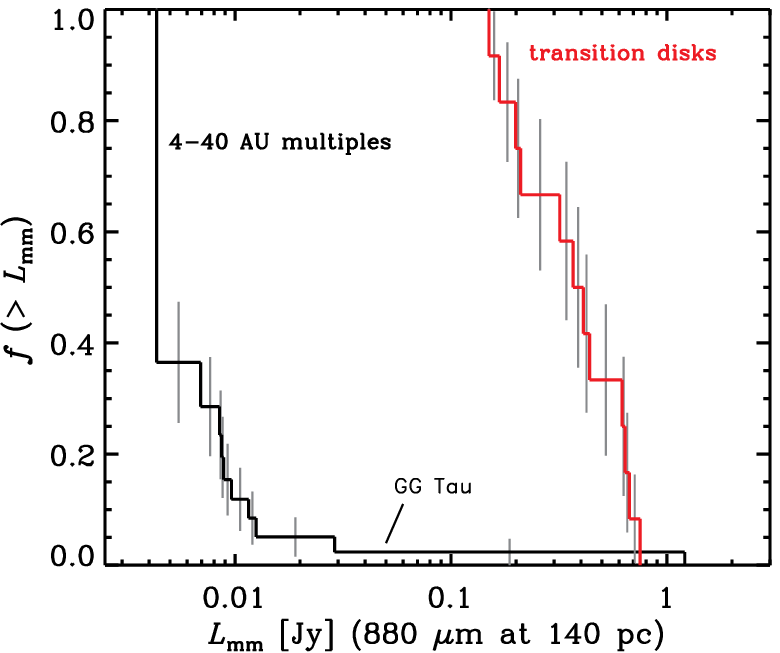}
\figcaption{({\it left}) The 880\,$\mu$m luminosity \citep[$L_{\rm
mm}$;][]{aw05,aw07b} as a function of the projected separation
\citep[$a$;][]{barsony03,kraus11} for multiple star systems in the Taurus and
Ophiuchus star-forming regions.  Gray points with arrows signify 3\,$\sigma$
upper limits on $L_{\rm mm}$.  The red lines mark the stellar companion
separations that would be expected to clear the transition disk cavities: their
vertical extents correspond to the range of eccentricities $e \sim 0.0$-0.9
\citep{artymowicz94,pichardo08}.  ({\it right}) A direct comparison of the
cumulative $L_{\rm mm}$ distributions for the transition disks (red) and the
multiples with projected separations of 4-40\,AU (black; the latter was
constructed with the Kaplan-Meier estimator to include 3\,$\sigma$ upper
limits).  The two populations have extremely different millimeter-wave
luminosity (disk mass) distributions: transition disks are typically a factor
of $\sim$30$\times$ brighter.  \label{companions}}
\end{figure*}

If the transition disk cavities were created by tidal interactions with stellar 
companions, their expected orbital separations would be $a \sim 0.3$-0.6$R_{\rm 
cav}$, for high eccentricities ($e \sim 0.9$) down to circular orbits.  The 
range of cavity sizes measured here would correspond to companion separations 
of $a \sim 4$-40\,AU.  The multiplicity fraction for similar primary star 
masses in this separation range is $\sim$10-20\%\ \citep{duquennoy91,kraus11}, 
which is similar to, but slightly below, the minimum frequency of 
millimeter-bright transition disks with large cavities we estimated in \S 
5.1.2.  Given how small those separations are projected on the sky, 
$\sim$0.04-0.20\arcsec, direct searches for companions have been limited by 
both resolution and contrast.  \citet{kraus11} have overcome such restrictions 
in Taurus, using a near-infrared aperture-masking interferometry technique 
coupled with the adaptive optics system on the Keck telescope.  For the four 
Taurus disks in our sample \citep[as well as RY Tau; see][]{isella10}, Kraus et 
al.~definitively rule out the presence of any companions with $q \ge 0.01$ 
($\ge$0.02 for DM Tau) in the separation ranges implied by $R_{\rm cav}$ 
\citep[see also][]{pott10}.  The same applies to the LkH$\alpha$ 330 and SR 21 
disks (A.~L.~Kraus 2011, private communication).  For {\it at least} half of 
the known transition disks with large cavities, there are no companions with 
$M_s \gtrsim 20$-30\,M$_{\rm jup}$ that could be responsible for the observed 
disk structures.  

Until similar measurements are available for the other transition disks, their 
multiplicity status will remain uncertain.  However, there are some stark 
differences between the transition disks and $a = 4$-40\,AU multiples.  Tidal 
interactions substantially diminish disk masses in multiples with this 
separation range \citep{jensen94,jensen96,osterloh95,aw05,cieza09}.  To 
illustrate that point, we compare the millimeter-wave luminosities ($L_{\rm 
mm}$, see \S 5.1.2) with the projected separations of the known multiples in 
Ophiuchus and Taurus \citep{barsony03,kraus11} in Figure \ref{companions}a.  
The transition disks are marked with red lines, representing the companion 
separation ranges implied by their cavity sizes.  More directly, Figure 
\ref{companions}b compares the cumulative $L_{\rm mm}$ distributions of the 
transition disks and the known 4-40\,AU multiples \citep[the latter includes 
upper limits and was constructed using the Kaplan-Meier product limit 
estimator; see][]{feigelson85}.  The two populations have remarkably different 
$L_{\rm mm}$ (or equivalently, $M_d$) distributions: the median transition disk 
is at least 30$\times$ brighter (more massive) than the 4-40\,AU multiples.  Of 
the 44 known multiples in that separation range, only one - GG Tau - is 
brighter than 30\,mJy at 880\,$\mu$m.  GG Tau is exceptionally rare, hosting a 
circumbinary ring not unlike the transition disk structures (it technically 
meets our definition of a transition disk in \S 5.1.1), but with an enormous 
$R_{cb} \approx 180$\,AU \citep{guilloteau99}.  In any case, the absence of 
millimeter-bright 4-40\,AU multiples is striking: it suggests that the specific 
type of transition disk studied here has not suffered from the same destructive 
interaction mechanism.

Some caution is warranted in over-interpreting the differences between these 
two populations, as selection biases, unknown orbital parameters ($e$, $i$), 
and perhaps uncertainties in the tidal truncation models \citep[e.g., 
see][]{beust05} should also be considered.  One notable distinction is related 
to the mass ratio.  There are stringent limits on $q$ ($\ge 0.01$-0.02) for at 
least half of the transition disk sample, but the 4-40\,AU multiples are 
preferentially equal-mass \citep[median $q \approx 0.7$;][]{kraus11}.  Such 
high mass ratios could impart more severe tidal perturbations to the 
circumstellar environment at an early evolutionary stage.  Since that seems 
unlikely for the transition disks, it is worthwhile to explore the effects of 
companions with much lower $q$ values.  Such low-mass companions that form in 
the disk around the primary - including giant planets and brown dwarfs - can 
modify the disk structure without dispersing too much of the total mass 
($\propto L_{\rm mm}$).  

Resonant torques induced by a low-mass companion set up spiral waves 
that repel local disk material \citep{goldreich78,goldreich80,lin79a,lin79b}.  
Above a mass ratio threshold, $q \gtrsim q_{\rm \, gap} \approx 40 \alpha 
(H/a)^2$ \citep[where $\alpha$ is the viscosity coefficient;][]{shakura73}, 
these tidal interactions open a gap in the disk 
\citep{lin86a,lin93,bryden99,crida06}.  In this scenario, the companion is 
located closer to the outer edge of the cavity than for the high-$q$ case.  A 
typical separation for a circular orbit might be $a \sim 0.8 R_{\rm cav}$, 
although the exact location depends on some complicated issues that influence 
the gap width \citep{winters03,crida06}.  For our model structures and $\alpha 
= 0.01$, the transition disks in this sample would need companions with $a \sim 
10$-60\,AU and mass ratios at least $q_{\rm \, gap} \approx 0.001$-0.03 to open 
gaps near their cavity edges.  Using the $M_{\ast}$ values in Table 
\ref{stars_table}, those limits translate to companion masses $M_s \gtrsim 
1$-70\,M$_{\rm jup}$ (the high end corresponding to larger $M_{\ast}$; note 
that $q_{\rm \, gap}$ and $M_s$ scale linearly with the assumed $\alpha$).  For 
the relevant cases, the mass ratios implied from current infrared contrast 
limits \citep{pott10,kraus11} exceed those needed to open gaps.  Although 
reassuring, there is little to be learned there, given the uncertainties on the 
disk viscosities ($\alpha$) and the luminosities of such low-mass companions 
\citep[e.g.,][]{marley07}.  

An individual gap is too narrow to be responsible for the observed properties 
of the transition disks.  However, the disk-companion interactions can also 
regulate the densities in the inner disk, interior to the companion orbit.  As 
in the case for a photoevaporative flow (\S 5.3.1), the gap effectively 
isolates the inner disk from its replenishment reservoir at larger radii.  
Without a viscous flow of material inwards, the inner disk material would drain 
onto the star on a relatively short timescale, producing a large and empty 
cavity.  Unlike the photoevaporation mechanism, some material from the outer
disk is expected to seep through a gap created by a low-mass companion in 
accretion streams, but only at decreased flow rates compared to a normal disk
\citep[e.g.,][]{artymowicz96,kley99,lubow99}.  \citet{lubow06} estimate that
the mass flow across such a gap is diminished by $\sim$75-90\%, depending on 
$q$: a more massive companion reduces the flow across the gap, leading to a 
lower-density inner disk.  A quantitative association between the companion 
mass, the accretion flow, and the depletion of the inner disk is not yet clear 
in simulations.  Some have suggested a companion with $q \approx 0.001$ is 
sufficient to deplete a disk cavity like those observed for the transition 
disks \citep{rice03,quillen04,varniere06}, while others have argued that 
higher masses are required \citep{crida07}.  

\begin{figure}
\epsscale{1.1}
\plotone{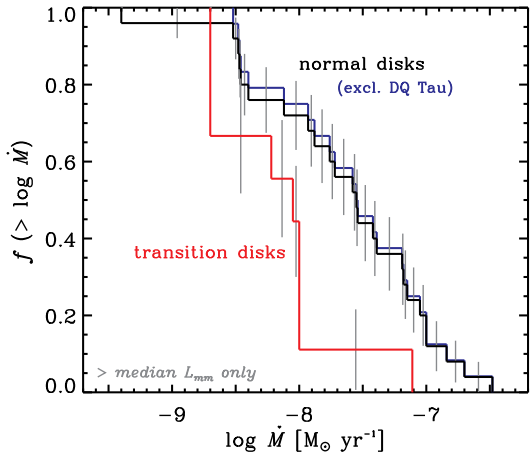}
\figcaption{The cumulative distribution of the accretion rates ($\dot{M}$) for
the normal disks in Tau and Oph that reside in the bright half of the
$L_{\rm mm}$ distribution (black, or blue excluding the spectroscopic binary DQ
Tau), compared with that for the transition disks with large cavities in those
same regions (red; including RY Tau).  The latter was constructed with the
Kaplan-Meier estimator to include the upper limit on $\dot{M}$ for the SR 21
disk.  We do not include the candidate transition disk RX J1633-2422 in either
distribution, although it has a very low $\dot{M}$ estimate 
\citep[see][]{cieza10}.  Accretion rates were compiled from \citet{hartmann98}, 
\citet{natta06}, and \citet{najita07}.  Of the 46 millimeter-bright disks 
considered, 7 (primarily in Oph) have no $\dot{M}$ measurements in the 
literature.  In this preliminary comparison, the transition disks typically 
have accretion rates $\sim$3$\times$ (0.5\,dex) lower than the normal disks.  
However, it is difficult to reach any definitive conclusions from such a 
comparison: the two samples are small and likely incomplete (and biased), and 
the $\dot{M}$ values have large uncertainties and were determined in different 
ways.  \label{fMdot}}
\end{figure}

\citet{najita07} suggested that transition disks exhibit low accretion rates 
compared to their normal counterparts with similar disk masses, consistent with 
a large reduction in the mass flow across a gap produced by a low-mass 
companion \citep[see also][]{alexander07,alexander09}.  To verify that result, 
we constructed the distribution of $\dot{M}$ for normal disks in Taurus and 
Ophiuchus that lie in the bright half of the $L_{\rm mm}$ distribution and 
compared it with the accretion rates for the transition disks with large 
cavities in those regions (including RY Tau; see Table \ref{stars_table}).  
Figure \ref{fMdot} shows the cumulative distributions of $\dot{M}$ for both of 
these populations.  A two-sided Kolmogorov-Smirnov test \citep[or its survival 
analysis equivalents if we include the $\dot{M}$ upper limit for SR 21; 
see][]{isobe86} indicates that the probability these $\dot{M}$ populations are 
drawn from the same parent distribution is low ($\le$1-2\%), with the 
transition disks having about $\sim$3$\times$ lower accretion rates on 
average.  That relative $\dot{M}$ decrease ($\sim$70\%) is similar to the 
expected reduction in flow rates across a gap opened by a $q \approx 0.001$ 
(i.e., $M_s \approx 1$\,M$_{\rm jup}$) companion \citep{lubow06}.  However, the 
samples are small, incomplete, and likely biased.  Moreover, the individual 
accretion rates have large uncertainties and are estimated from different 
diagnostics.  So, although this preliminary comparison \citep[and that 
by][]{najita07} provides a tantalizing hint of reduced mass flow rates for 
transition disks, we caution that a more rigorous and homogeneous set of 
comparisons is warranted before any conclusions are drawn.

% FIG 12

Regardless of those comparisons, the $\dot{M}$ values for the transition disks 
are still high enough to be seemingly inconsistent with the low dust densities 
we infer inside their cavities.  One way to reconcile those 
measurements is to include additional low-mass companions inside the observed 
disk cavity.  \citet{zhu11} have suggested that multi-planet systems can 
dynamically clear large cavities while also rapidly transporting gas across the 
large distances between the remnant outer disk and the star.  Alternatively, a 
single low-mass companion might be responsible for creating a cavity that is 
depleted in dust, but not in gas.  Any companion that can open a gap will 
generate wakes just outside its orbit, creating a pressure bump that may act 
like a dust filter on the flow across the gap \citep{rice06,paardekooper06}.  
If that filter removes a substantial fraction of the dust from the flow, the 
inner disk can be preferentially depleted of dust (producing the thermal 
deficits we observe) while still maintaining high gas accretion rates.  

The tidal interactions between disk material and low-mass companions offer a 
compelling explanation for the transition disk structures.  However, the 
observations are only starting to provide some tangible constraints that can be 
incorporated into the numerical simulations.  As the data improve - both for 
the disk structures and companion searches - we should be able to make a more 
nuanced assessment of whether the mechanism responsible for transition disk 
cavities is tied directly to the formation and early evolution of low-mass 
brown dwarfs and/or exoplanets.

\subsection{Implications for Planet Formation}

To summarize the discussion in \S 5.2, we have argued that the dust-depleted 
cavities in the specific type of transition disks studied here - those with 
large $L_{\rm mm}$ and $R_{\rm cav}$ - were not likely produced by 
photoevaporative winds or tidal truncation by stellar-mass companions.  The 
possibility that the transition disk cavities are the manifestation of 
substantial particle growth in large, low-ionization dead zones remains, but 
requires a more detailed modeling effort to determine if some significant 
challenges could be mitigated.  While these mechanisms all play non-trivial 
roles in disk evolution, the weight of the observational evidence for the kind 
of transition disks studied here suggests that the cavities are likely the 
byproduct of dynamical interactions with long-period low-mass companions - 
either brown dwarfs or giant planets.  If that is the case, it is of great 
interest to understand how those companions can form at (or be moved to) such 
large separations from their host stars ($a \sim 10$-60\,AU) on such a 
relatively short timescale (at most $\sim$1-2\,Myr).  

Giant planets or low-mass brown dwarf companions could be formed relatively 
rapidly from a fragmentation process in a gravitationally unstable disk 
\citep[e.g.,][]{boss97,boss98,rice03b}.  Moreover, this disk instability 
mechanism seems to naturally favor formation at wide separations from the 
primary \citep{boss06,boley09,rafikov09,clarke09,dodson-robinson09}.  Of 
course, it is not clear that the transition disks initially had sufficiently 
high densities (and efficient cooling) to be unstable, and in most cases (8/12 
in this sample) the orbits of any companions would be small enough ($a \le 
35$\,AU) that the prospects for fragmentation are debatable \citep[but 
see][]{boss09}.  If the companions are not too massive, the more standard core 
accretion mechanism for giant planet formation may be applicable.  In that 
scenario, the collisional agglomeration of solids must produce a massive core 
to accrete a massive atmosphere before the gas disk disperses 
\citep[e.g.,][]{pollack96,hubickyj05}.  Traditionally, that process has been 
considered too inefficient to make long-period planets ($a \gtrsim 10$\,AU) 
{\it in situ} within the disk lifetime.  However, \citet{rafikov11} has argued 
that an improved treatment of planetesimal accretion shows expedited core 
growth at low surface densities, thereby enabling the capture of a massive 
atmosphere in the outer disk.  His estimate of the $\ge$0.1\,g cm$^{-2}$ of 
solids needed for that process is strikingly similar to (or below) the $\Sigma$ 
values we derive near $R_{\rm cav}$ (see Figure \ref{sigma}).  Others have 
suggested that long-period companions could be produced relatively easily by 
the core accretion mechnism in the inner disk, and then subsequently moved 
outward via planet-planet scattering \citep{scharf09,veras09} or a resonant 
migration process \citep{masset01,crida09}.

\subsubsection{Future Prospects}

At this stage, it is not clear that the properties of transition disks can help 
differentiate between these planet formation mechanisms - even if it can be 
confirmed that their structures are produced by planetary-mass companions.  
However, a future census of companions embedded in such disks that determines 
the joint distribution of \{$a$, $M_s$\} would help guide that debate.  In that 
context, the highest priority should be placed on continuing the high-contrast 
and high-resolution infrared searches for low-mass companions inside the known 
transition disk cavities, at the very least to rule out the presence of their 
more massive (stellar) counterparts \citep[along the lines of][]{kraus11}.  The 
identification of any young exoplanet candidates still embedded in their natal 
disk material would be a landmark discovery for both planet formation and disk 
evolution.  Moreover, those companions would provide youthful touchstones that 
could help facilitate a better understanding of their much older counterparts, 
including the Fomalhaut and HR 8799 planets that orbit just inside their own 
remnant debris disk rings \citep{kalas08,marois08,marois10}.  

While those companion searches continue, there are two issues related to the 
planet formation process that can be better addressed with observations of the 
transition disks themselves.  First, the Atacama Large Millimeter Array (ALMA) 
project starts operations soon and will routinely achieve extraordinary 
millimeter continuum sensitivity with sub-arcsecond resolution.  Those 
capabilities will permit the first large, unbiased, and direct searches for 
large ($R_{\rm cav} > 10$\,AU) dust-depleted cavities in nearby disks, and 
should provide a more definitive assessment of transition disk properties 
across the entire disk mass distribution (see \S 5.1.2).  These imaging surveys 
will provide a first look at the fraction of disks that have formed long-period 
giant planets within $\sim$1-2\,Myr, and can be compared with direct imaging 
exoplanet searches around older stars in the solar neighborhood.  As those 
surveys are building up statistics, a significant effort should be invested in 
developing our understanding of the structure and contents of the material 
inside the transition disk cavities.  With a focus on individual sources, ALMA 
will be sensitive enough to provide stringent constraints on the depletion of 
mm-sized grains and molecular gas in these cavities, and could potentially 
detect faint emission signals that can be related to the inner disk mass.  The 
residual 880\,$\mu$m signal in the LkCa 15 disk cavity is a tantalizing hint of 
the prospects for such studies.  Coupled with estimates of the cavity size and 
accretion rate, that information can be used in concert with numerical 
simulations to help infer the mass and orbital period of the planetary 
companion perturbing the disk structure.  While the present study is just one 
of the first steps, ultimately high angular resolution radio mapping of disk 
structures can serve as a new tool to guide searches for young exoplanets, as 
well as to characterize some of their fundamental properties.

\section{Summary}

We have presented a sample of new and archival SMA observations of the 
880\,$\mu$m continuum emission from protoplanetary transition disks at 
resolution scales down to $\sim$0\farcs3 (20-40\,AU).  Using two-dimensional 
Monte Carlo radiative transfer calculations and a parametric formulation for 
the disk densities, we simultaneously modeled these SMA visibilities and the 
broadband SEDs for each disk in a homogeneous framework.  The data and those 
modeling results were then synthesized to discuss the generic properties of 
transition disks from a primarily millimeter-wave perspective, as well as to 
assess the roles of various potential disk evolution mechanisms.  Our key 
results include: 

\begin{enumerate}
\item In all cases, the SMA data resolve large ($R_{\rm cav} = 15$-73\,AU), 
dust-depleted cavities at the disk centers.  Unlike estimates based on models 
of the unresolved infrared spectra, these {\it direct} imaging measurements 
constrain the cavity sizes with little ambiguity: $R_{\rm cav}$ uncertainties 
are typically only $\sim$10\%\ (or up to 15-20\%\ if a range of surface density 
profiles are considered).  

\item Some dust is still present inside these disk cavities, although its mass 
and spatial distribution are poorly constrained.  With respect to continuous 
density models, we estimate from the SMA data that the densities of mm-sized 
particles are $\ge$10-100$\times$ lower inside the cavities.  From the infrared 
SED, we find that $\mu$m-sized particles are depleted by larger factors, 
typically 4-6 orders of magnitude.  Despite these low densities, the dust 
inside the cavities can have profound effects on the infrared spectrum and the 
structures we infer at larger radii.

\item Outside the cavities, the transition disk structures appear to be similar 
to disks with continuous density distributions: they have the same ranges of 
characteristic radii ($R_c$) and total masses ($M_d$).  In their mass 
reservoirs at large radii, the surface densities for the transition disks are 
slightly higher than normal disks with similar millimeter luminosities.  

\item Our sample is biased toward millimeter-bright disks with {\it large} 
($R_{\rm cav} \gtrsim 15$\,AU) dust-depleted cavities.  However, we find that 
this specific type of transition disk is surprisingly common among the 
Taurus/Ophiuchus disks with large millimeter luminosities (or disk masses): 
they comprise {\it at least} 20\%\ of the disks in the upper half of the 
$L_{\rm mm}$ (or $M_d$) distribution, and $>$25\%\ in the upper quartile.  
Imaging surveys of other millimeter-bright disks rule out large-cavity 
transition disk fractions in excess of 70\%, but there is no such information 
available for smaller cavities or fainter disks.  The transition disk fractions 
derived from direct images of cavities are higher than those estimated from 
infrared colors because the latter technique systematically misses cases with 
excess emission from a small amount of dust inside the cavity.

\item The large cavity sizes, relatively high accretion rates, and low X-ray 
luminosities suggest that the transition disk cavities in this sample were not 
cleared by an internal photoevaporation process (as was already argued by the 
developers of those dispersal models).  

\item If the diminished signatures of dust grains in the cavities are to be 
explained by particle growth alone, the transition disk structures need to 
contain very large MRI-inactive dead zones, have experienced growth to at least 
meter-sized solids, and harbor only very limited reservoirs of small dust 
grains in their inner disks (from collision fragments or dust coupled to a 
layered accretion flow).  While those stringent conditions make this mechanism 
for cavity formation seem challenging, a real assessment of the role of 
particle growth in producing transition disk structures cannot be made until 
more detailed models are explored.

\item In principle, dynamical interactions with stellar companions on $a 
\approx 4$-40\,AU orbits would be capable of opening the disk 
cavities we measure.  However, recent infrared searches definitively rule out 
any such companions with mass ratios $q \gtrsim 0.01$ for half of this sample.  
Moreover, the transition disks studied here have substantially higher 
(typically $\ge$30$\times$) millimeter-wave luminosities (or disk masses) than 
the disks in multiple star systems with 4-40\,AU separations, suggesting that 
they did not suffer the same level of mass loss from tidal perturbations.

\item We find the best overall agreement with the data if these transition disk 
cavities are opened by tidal interactions with low mass ratio ($q \gtrsim 
0.001$) companions - either low-mass brown dwarfs or giant planets.  Such a 
companion can open a gap and substantially deplete material interior to its 
orbit without disrupting the outer disk.  We demonstrate that the typical 
accretion rates for these transition disks are decreased (by $\sim$70\%) 
compared to their high-$L_{\rm mm}$ normal disk counterparts, commensurate with 
the reduced mass flow that is expected to cross the orbit of a low-mass 
companion (however, a more in-depth analysis with larger samples is 
warranted).  In the near future, radio imaging studies of transition disk 
structures could provide a new and unique method for identifying and 
characterizing very young ($\sim$1\,Myr) low-mass companions -- including giant 
planets on long-period orbits -- still embedded in their natal disk material. 

\end{enumerate}

\acknowledgments We thank Karin \"{O}berg, Sijme-Jan Paardekooper, and 
Aur{\'{e}}lien Crida for beneficial conversations, and a referee for a thorough 
review that helped clarify the text.  We are especially grateful to Adam Kraus, 
who kindly shared his results prior to publication.  The SMA is a joint project 
between the Smithsonian Astrophysical Observatory and the Academia Sinica 
Institute of Astronomy and Astrophysics and is funded by the Smithsonian 
Institution and the Academia Sinica.

\clearpage

\appendix

\section{Comments on Individual Sources}

This appendix highlights some aspects of the modeling results for individual
disks, and compares with previous efforts to characterize their
disk cavities whenever available.  Table \ref{seds_table} is a compilation of 
the original sources used to construct the SEDs displayed in Figures \ref{results1} and \ref{results2}.

\paragraph{MWC 758 --} This Herbig Ae star is relatively isolated, located east 
of the main Taurus-Auriga clouds.  Without a firm cluster association, its 
distance is uncertain but assumed here to be the nominal {\it Hipparcos} value, 
$d = 200$\,pc \citep{perryman97}.  We adopt the A8 spectral classification 
advocated by \citet{beskrovnaya99}, but note that others have utilized an 
earlier type \citep[A5;][]{vandenancker98}.  The SMA observations from the E 
configuration are the same as presented by \citet{isella10b}, although they 
have been re-calibrated with an amplitude scale that amounted 
to a $\sim$30\%\ decrease in flux levels.  That modification was required to 
reconcile those data and our new SMA observations with short antenna spacings 
(C configuration); the combined data are now in excellent agreement with 
single-dish photometry measurements.  The 880\,$\mu$m visibility null calls for 
the largest cavity in the sample, $R_{\rm cav} = 73$\,AU, despite the bright 
infrared excess.  The standard infrared signatures of a transition disk are 
heavily muted, washed out by only a small amount of dust inside the cavity.  
Although that dust has densities that are comparable to the other disks in the 
sample (and only modest rim and cavity heights), it produces a significantly 
brighter infrared spectrum for two reasons: (1) the central star is 
extraordinarily hot and luminous, substantially heating more dust, and (2) the 
scale heights are large, increasing the irradiated surface area and therefore 
the emitted flux from the disk atmosphere.  

\citet{isella10b} also modeled the MWC 758 disk structure with some of these 
SMA data (and supplementary 3\,mm observations), choosing not to modify the 
accretion disk model to include a cavity (i.e., $\Sigma = \Sigma_g$ at all 
radii in their work) but to allow the gradient of the viscosity profile 
($\gamma$) to vary freely.  They can reproduce the data with a large negative 
viscosity gradient, $\gamma \approx -4$, which results in a surface density 
profile that increases like $R^4$ inside $\sim$70\,AU, peaks near 100\,AU, and 
then drops off like $1/\exp{(R^6)}$ at larger radii.  Roughly speaking, that 
density distribution is similar to the narrow ring employed here (see Figure 
\ref{sigma}), even peaking at essentially the same location and surface 
density.  Not unexpectedly, the Isella et al.~model requires an additional dust 
component near the star to account for the infrared excess (which their $R^4$ 
density profile alone vastly under-produces), similar to the inner disk 
material utilized here.  Finally, the sharp density cutoff at large radii that 
is inherent to models with negative $\gamma$ values means that the Isella et 
al.~model has effectively no mass outside $R \approx 150$\,AU.  On the 
contrary, our model has gas densities in excess of 10$^{-5}$\,g cm$^{-2}$ out 
to $R \approx 400$\,AU, despite the small characteristic radius 
(because $\gamma = 1$).  We expect that this difference in the effective disk 
size would help explain the strong positive residuals at large radii noted by 
Isella et al.~when applying their dust model to the CO emission.  

\paragraph{SAO 206462 --} Commonly referred to as HD 135344 B, this star is 
also isolated, perhaps associated with the outskirts of the Sco-OB2 
association.  We adopt the distance estimate of \citet{vanboekel05}, $d = 
140$\,pc, and the F4 spectral type determined by \citet{dunkin97}.  The SMA 
data were first presented by \citet{brown09}, and only include observations in 
the V configuration with few short baselines.  Our models imply a large cavity, 
$R_{\rm cav} = 46$\,AU, with a small amount of dust near the star.  The 
super-heated inner rim near the dust sublimation radius is responsible for the 
substantial near-infrared bump in the SED.  To minimize the strength of the 
10\,$\mu$m silicate feature, comparatively large ($s_{\rm max} = 3$\,$\mu$m) 
dust grains were used to populate the inner disk.  Our model tends to 
over-predict the flux densities at wavelengths longward of 1\,mm, a not 
uncommon feature due to our decision to fix the grain size distribution in the 
midplane of the outer disk (see \S 3.2).  The issue might also be alleviated if 
new data with short antenna spacings become available, enabling a more robust 
disk mass estimate.  The SMA image of the SAO 206462 disk is notably 
asymmetric, with enhanced emission to the southeast and a pronounced deficit to 
the north.  But the imaged residuals indicate that those asymmetries are only 
marginally significant: given the challenges of observing such a southern 
target, those features are not inconsistent with noise \citep[but note that a 
similar pattern is present in the scattered light images presented 
by][]{grady09}.  

\citet{brown07} also used {\tt RADMC} calculations to reproduce the SED alone, 
and later combined with these SMA data \citep{brown09}, but with a different 
density model (a power-law $\Sigma$ profile, truncated at the edges).  They 
estimated roughly the same cavity radius, 39-45\,AU, consistent within the 
expected $\sim$10\%\ mutual uncertainties.  Although their lower $R_{\rm cav}$ 
value was determined from the SMA data, the difference with the cavity size 
derived here is likely due to small offsets in the adopted disk centroids and 
viewing geometries.  Regarding the latter, we utilize the inclination 
determined by \citet{lyo11} from CO spectral line images, in good agreement 
with the scattered light constraints of \citet{grady09}.  Those scattered light 
images suggest a substantially settled vertical distribution of dust with an 
effective $\psi = -0.3$, unlike the more common $\psi = 0.15$ used here.  Part 
of that difference is related to shadowing from the inner rim/disk and cavity 
wall, which can effectively steepen the scattered light surface brightness 
profile.  But there is so little radial dynamic range available from this disk 
that it should be possible to find anti-flaring models that are also 
commensurate with the SMA data.

\paragraph{LkH$\alpha$ 330 --} Associated with the Perseus molecular clouds 
\citep[$d \approx 250$\,pc; see][]{enoch06}, this young star was classified as 
spectral type G3 by \citet{coku79}.  The SMA data in the V configuration were 
first presented by \citet{brown08}; new SMA data in the C configuration are 
also included here.  Note that there is considerable disagreement between the 
{\it IRAS} and {\it Spitzer} data in the far-infrared, perhaps due to molecular 
cloud or cirrus contamination in the former.  When re-calibrating the 
long-baseline SMA data used by \citet{brown08,brown09}, it was noted that the 
observations of 2006 November 11 had such poor phase stability that they could 
not be used.  With their removal and the addition of new data with short 
antenna spacings, a much more symmetric and sensitive map of the 880\,$\mu$m 
continuum emission is available.  

We measure a large cavity, $R_{\rm cav} = 68$\,AU, $\sim$35-60\%\ larger than 
estimated by \citet{brown08,brown09} from (some of) the same data.  Most of 
that discrepancy is caused by an inaccurate estimate of the 880\,$\mu$m 
visibility null location, due to the poor quality of one of the SMA tracks in 
the V configuration.  Using the CO $J$=3$-$2 spectral line images from the new 
data in the C configuration, we also use an updated viewing geometry (with a 
significant difference in the major axis PA; the full CO datasets will be 
presented elsewhere).  The faint 10\,$\mu$m silicate feature was difficult to 
reproduce with the typical grain size index ($p = 3.5$), but we found some 
success using a more top-heavy distribution ($p = 2.5$) for the dust inside the 
cavity.  To reproduce the relatively bright infrared excess with that grain 
population, a substantial increase in the inner rim height was required (the 
dust size distribution parameters and rim height are strongly degenerate; see 
\S 3.5).

\paragraph{SR 21 --} This young star with spectral type G3 is still embedded in 
the L1688 dark cloud in the nearby Ophiuchus star-forming region, $d \approx 
125$\,pc \citep[e.g.,][]{lombardi08,loinard08}.  The extinction estimate of 
\citet{prato03}, $A_V \sim 9$, is too large to be appropriate for both the 
optical photometry of \citet{vrba93} and the near-infrared magnitudes in the 
2MASS catalog \citep{cutri03}.  Assuming variability is not responsible for 
that discrepancy, we adopt a lower extinction ($A_V = 6.3$) and therefore a 
substantially lower $L_{\ast}$ compared to \citet{brown07,brown09}.  The SMA 
data are the same as described by \citet{andrews09}, although the 880\,$\mu$m 
continuum map has been synthesized with slightly improved resolution.  The 
880\,$\mu$m continuum emission is distributed in a nearly face-on ring, with 
$R_{\rm cav} = 36$\,AU.  The narrow ring width ($R_c$ is only 15\,AU) makes it 
difficult to determine the viewing geometry, and there is no uncontaminated CO 
emission or scattered light image available to facilitate an independent 
measurement.  The lack of a clear silicate feature in the mid-infrared leads us 
to adopt a large $s_{\rm max}$ (10\,$\mu$m) for the dust inside the cavity.  
Despite the different modeling setups, the disk structure parameters inferred 
here are essentially the same as those determined by \citet{andrews09}.  With 
(some of) the same SMA data, \citet{brown09} also find a comparable cavity size 
(33\,AU).  We find positive 3\,$\sigma$ residuals for this model, suggesting 
excess emission to the southeast and near the stellar position.  Those 
discrepancies lead to a poor prediction of the secondary visibility null (see 
Figure \ref{results1}), which may indicate that there is a small amount of 
$\sim$mm-sized dust emitting from inside the cavity.

\paragraph{UX Tau A --} The primary star of a wide multiple system in Taurus 
\citep[$d \approx 140$\,pc; see][]{torres09}, UX Tau A is the only component 
with evidence for remnant disk material \citep[e.g.,][]{white01,mccabe06}.  No 
880\,$\mu$m emission above 4.5\,mJy (3\,$\sigma$) is detected around either the 
UX Tau B or C components \citep[in agreement with][]{jensen03}.  The SMA data 
presented here are the first resolved observations of the UX Tau A disk.  This 
emission is quite compact: we use a small characteristic radius ($R_c = 
20$\,AU), presumably produced by the truncation of the outer disk by UX Tau C 
(located $\sim$2\farcs5 to the west).  Unlike the compact continuum, we find a 
bright, resolved CO $J$=3$-$2 disk orbiting UX Tau A, providing the first 
direct estimates of the disk inclination and position angle.  As discussed in 
\S 5.1, the 880\,$\mu$m image shows a resolved cavity that is substantially 
smaller, $R_{\rm cav} = 25$\,AU, than initial estimates from the SED alone 
\citep[56 and 71\,AU by][respectively]{espaillat07,espaillat10}.  Like 
LkH$\alpha$\,330, we assume a top-heavy dust size distribution ($p = 2.5$) in 
the cavity to avoid over-producing the 10\,$\mu$m silicate feature.

\paragraph{SR 24 S --} Embedded in the L1688 dark cloud in Ophiuchus, SR 24 is 
a hierarchical triple system with a close binary \citep[0\farcs2;][]{simon95} 
located $\sim$5\arcsec\ north of a spectral type K2 primary (SR 24 S).  The SR 
24 N binary exhibits some thermal excess presumably associated with a low-mass 
circumbinary disk, although it is not detected at millimeter wavelengths 
\citep[despite evidence for associated CO line emission;][]{aw05a}.  Due to 
potential confusion with the SR 24 N component, we did not use a {\it Spitzer} 
IRS observation of this system (although one apparently exists).  The 
880\,$\mu$m image of the SR 24 S disk shows an elongated ring with a 
significant brightness asymmetry along the major axis (brighter to the north).  
We measure a moderate cavity size, $R_{\rm cav} = 29$\,AU, essentially the same 
as our previous effort \citep[32\,AU;][]{andrews10b}.  Naturally, the model 
does not reproduce the observed asymmetry, instead over-predicting the emission 
to the southwest (although marginally, based on the imaged residuals).  In 
general, the fit quality here is superior to our initial work, thanks to the 
modification of the inner rim/disk structure.  However, without a detailed 
infrared spectrum the dust properties and structure of the cavity are 
uncertain.  Unfortunately, the asymmetry of the narrow emission ring makes a 
clear viewing geometry estimate difficult, and the CO emission from SR 24 S is 
contaminated by local cloud material.  Substantial improvements in the 
structure constraints are feasible with a more complete SED, if the disk 
inclination can be measured independently (e.g., with a tracer like $^{13}$CO).

\paragraph{DoAr 44 --} Sometimes referred to as ROXs 44, this young star is 
located in the densely-populated L1688 dark cloud in Ophiuchus.  A preliminary 
analysis of the DoAr 44 disk structure was presented by \citet{andrews09}, 
using the same SMA data.  As in that work, we find a moderate cavity size for 
the DoAr 44 disk, $R_{\rm cav} = 30$\,AU.  However, our new model is in much 
better agreement with the infrared SED, thanks to the inclusion of an inner rim 
and cavity wall in the model structure.  To match the relatively faint 
far-infrared emission, the dust in the outer disk is substantially concentrated 
toward the midplane.  Unfortunately, there are no independent constraints on 
the disk inclination or orientation; the adopted inclination could be uncertain 
by $\sim$20\degr.  The small extent of the 880\,$\mu$m emission suggests that 
the disk is quite small (the ring is narrow), with $R_c = 25$\,AU.  While the 
western peak of the 880\,$\mu$m emission appears slightly brighter, the model 
residuals show that the asymmetry is not significant.  \citet{espaillat10} 
modeled the DoAr 44 SED and derived a slightly larger cavity size (36\,AU), but 
within the uncertainties.

\paragraph{LkCa 15 --} Located in the Taurus star-forming region, this 
transition disk has been studied extensively since its cavity was first 
resolved at 1.3\,mm by \citet{pietu06}.  Because of the large spatial extent 
and prominent depression in the 880\,$\mu$m emission, these new SMA 
observations make for a spectacular example of a resolved transition disk 
image.  The cavity size derived here, $R_{\rm cav} = 50$\,AU, is in excellent 
agreement with the initial measurement (46\,AU) by \citet{pietu06}.  The SED 
modeling by \citet{espaillat10} argues for a slightly larger cavity (58\,AU), 
although that would be difficult to reconcile with the well-defined 880\,$\mu$m 
visibility null (see Figure \ref{results2}).  The SEDs for the LkCa 15 and DoAr 
44 disks are quite similar, with both calling for relatively settled vertical 
dust distributions.  An examination of the modeling results in this case 
reveals a striking discrepancy with the data: the model predicts far too little
millimeter emission inside the cavity (leaving 6\,$\sigma$ residuals at the
stellar position).  The integrated residual flux density is small, 
$\sim$10\,mJy, but potentially important because it can provide a more direct
estimate for the dust mass inside the cavity.  For the sake of homogeneity with
the rest of this sample, we postpone a more detailed modeling of the LkCa 15
disk cavity until higher resolution images are available.

\paragraph{RX J1615-3255 --} Perhaps less well-known than the other disks in 
this sample, this system has been kinematically tied to the Lupus association, 
with $d \approx 185$\,pc \citep{makarov07}.  Although it had been identified as 
a weak-line T Tauri star \citep{henize76,krautter97}, infrared observations by 
{\it Spitzer} have only recently claimed RX J1615-3255 as a transition disk 
candidate \citep{merin10}.  The new SMA data presented here represent the first 
spatially resolved images of thermal emission from this disk.  A bright, 
rotating CO $J$=3$-$2 emission structure is also detected, providing a 
preliminary estimate of the viewing geometry.  Although much of the outer disk 
emission is spatially filtered in the synthesized 880\,$\mu$m maps shown in 
Figures \ref{images} and \ref{results2}, the visibilities clearly indicate that 
RX J1615-3255 hosts a rather large disk, with $R_c = 115$\,AU (in good 
agreement with the extent of the CO emission).  A particularly low-density 
cavity is present out to $R_{\rm cav} = 30$\,AU.  In fact, we were forced to 
empty the inner disk inside a radius of 0.5\,AU to avoid producing a 
near-infrared emission signature above photospheric levels (therefore, no inner 
rim was employed in this model; see Table \ref{structure_table}).  To reproduce 
the comparatively faint infrared excess beyond the dip in the SED, the scale 
heights were decreased to values consistent with substantial dust settling.  
The resulting low dust temperatures, coupled with the large disk size, lead to 
a relatively high estimate for the total disk mass ($\sim$0.13\,M$_{\odot}$, 
almost 12\%\, of the stellar mass).

\paragraph{GM Aur --} This star in the Taurus star-forming region is perhaps 
the canonical example of a transition disk, with many detailed discussions of 
the implications of its distinctive infrared SED 
\citep[e.g.,][]{strom89,skrutskie90,koerner93,chiang99,rice03}.  Using some of 
the same SMA data, \citet{hughes09} first resolved the central dust cavity.  We 
include new 880\,$\mu$m SMA observations here, with short (C) and intermediate 
(E) antenna spacings that provide a substantially improved Fourier sampling 
near the visibility null.  The cavity size derived here, $R_{\rm cav} = 
28$\,AU, is larger than the preferred \citet{hughes09} value (20\,AU) because 
the new SMA observations provide a more accurate location for the 880\,$\mu$m 
visibility null (at a slightly shorter deprojected baseline).  However, the 
alternative disk model discussed by \citet{hughes09} employed a larger cavity 
(26\,AU) that is in good agreement with our model.  The long-baseline SMA data 
is quite noisy in this case; improved sensitivity would help provide a more 
accurate cavity size estimate.  Like the \citet{hughes09} results, our model 
successfully matches a high resolution 1.3\,mm dataset from the Plateau de Bure 
interferometer.  The GM Aur disk is also found to be quite large, with a 
characteristic size $R_c = 120$\,AU (although most of that emission has been 
filtered out in the images displayed here; see Table \ref{image_table}).

\paragraph{DM Tau --} Although the gas disk around this cool, young star in the 
Taurus clouds has been studied extensively 
\citep[e.g.,][]{guilloteau98,dartois03,pietu07}, our new SMA observations 
permit the first sub-arcsecond resolution examination of the 880\,$\mu$m dust 
continuum emission.  Based on the shape of its {\it Spitzer} IRS spectrum, 
\citet{calvet05} suggested that the DM Tau disk has a relatively small cavity 
of radius $\sim$3\,AU.  Since such a small cavity cannot be resolved on the 
longest SMA baselines, it was a surprise to measure a clear 880\,$\mu$m 
emission ring and visibility null indicative of a much larger dust-depleted 
cavity, $R_{\rm cav} = 19$\,AU.  Reconciling this large cavity size in the SMA 
data with the morphology of the IRS spectrum was a challenge, and our current 
model should be considered preliminary until the latter data is better 
matched.  To account for the absence of an infrared excess shortward of 
$\sim$8\,$\mu$m, we were forced to remove all dust inside a radius of 1\,AU
(there is no inner rim in this case; see Table \ref{structure_table}).  Like 
the cases of GM Aur and LkCa 15, we find that the DM Tau disk is rather large, 
with $R_c = 135$\,AU, in good agreement with the extended CO emission studied 
elsewhere.

\paragraph{WSB 60 --} The dust-depleted cavity in the disk around this very 
cool star (spectral type M4) in the core of the Ophiuchus star-forming region 
was first discovered by \citet{andrews09}.  Compared to that initial study, we 
find a slightly smaller cavity here ($R_{\rm cav} = 15$\,AU, compared to 
20\,AU), partly related to our choice to fix the surface density gradient to 
$\gamma = 1$ in the new models.  The cavity is just barely resolved with our 
SMA observations, and it shows no obvious signatures of dust depletion at 
infrared wavelengths.  The latter point drives us to infer the smallest dust 
depletion factor inside the cavity for this sample, corresponding to only a 
factor of $\sim$50$\times$ lower densities in the inner disk than the nominal 
continuous model.  Considering the ``normal" appearance of the infrared SED, it 
is natural to speculate that there are potentially many other such disks that 
have similarly large cavities but will only be identified when high resolution 
radio images are available.

% TABLE A
\begin{deluxetable}{lcccccc}
\tablecolumns{7}
\tablewidth{0pt}
\tabletypesize{\small}
\tablecaption{Sources for SED Data\label{seds_table}}
\tablehead{
\colhead{Name} & \colhead{} & \colhead{optical} & \colhead{} & \colhead{infrared} & \colhead{{\it Spitzer} IRS} & \colhead{(sub)-mm} \\
\colhead{(1)} & \colhead{} & \colhead{(2)} & \colhead{} & \colhead{(3)} & \colhead{(4)} & \colhead{(5)}} 
\startdata
MWC 758         & \hspace{0.5cm} & 1    & & 1, 2, 3     & 1       & 1, 2 \\
SAO 206462      & \hspace{0.5cm} & 2    & & 1, 4, 5     & 2       & 3, 4, 5 \\
LkH$\alpha$ 330 & \hspace{0.5cm} & 3    & & 1, 3, 6, 7  & 3       & 5, 6, 7, 8 \\
SR 21           & \hspace{0.5cm} & 4    & & 1, 3, 6     & 4       & 9, 10, 11 \\
UX Tau A        & \hspace{0.5cm} & 5    & & 1, 3, 8, 9  & 5       & 12, 13 \\
SR 24 S         & \hspace{0.5cm} & 6, 7 & & 1, 6, 9, 10 & \nodata & 9, 10, 11 \\
DoAr 44         & \hspace{0.5cm} & 8    & & 1, 6, 11    & 4       & 9, 14 \\
LkCa 15         & \hspace{0.5cm} & 9    & & 1, 12       & 5       & 12, 15, 16, 17, 18 \\
RX J1615-3255   & \hspace{0.5cm} & 10   & & 1, 3, 13    & 6       & 19 \\
GM Aur          & \hspace{0.5cm} & 11   & & 1, 3, 8     & 7       & 12, 17, 20, 21, 22, 23, 24 \\
DM Tau          & \hspace{0.5cm} & 5    & & 1, 3, 8     & 7       & 17, 25, 26 \\
WSB 60          & \hspace{0.5cm} & 7    & \hspace{0.3cm} & 1, 6, 11    & 4       & 9 \\
\enddata
\tablecomments{Col.~(1): Name of host star.  Col~(2): Optical references: 
[1] - \citet{vieira03}, 
[2] - \citet{coulson95},
[3] - \citet{fernandez96},
[4] - \citet{vrba93},
[5] - \citet{kenyon95},
[6] - \citet{herbig88},
[7] - \citet{wilking05},
[8] - \citet{herbst94},
[9] - \citet{bouvier93},
[10] - \citet{makarov07},
[11] - \citet{bouvier95}
Col.~(3): Infrared references:  
[1] - 2MASS \citep{cutri03},
[2] - \citet{malfait98},
[3] - {\it IRAS} \citep{beichman88},
[4] - \citet{coulson95},
[5] - \citet{walker88},
[6] - \citet{evans03},
[7] - \citet{rebull07},
[8] - \citet{luhman10},
[9] - \citet{mccabe06},
[10] - \citet{jensen97},
[11] - \citet{padgett08},
[12] - \citet{rebull10},
[13] - \citet{padgett06},
Col.~(4): {\it Spitzer} IRS spectrum references:   
[1] - the {\it Spitzer} archive, reduced as in \citet{mcclure10},
[2] - \citet{brown07},
[3] - \citet{brown08},
[4] - \citet{mcclure10},
[5] - \citet{espaillat07},
[6] - \citet{evans03},
[7] - \citet{calvet05}
Col.~(5): Millimeter references: 
[1] - \citet{difrancesco08},
[2] - \citet{chapillon08},
[3] - \citet{coulson95},
[4] - \citet{sylvester96},
[5] - \citet{brown09},
[6] - \citet{brown08},
[7] - \citet{osterloh95},
[8] - \citet{enoch06},
[9] - \citet{aw07b},
[10] - \citet{andre94},
[11] - \citet{patience08},
[12] - \citet{aw05},
[13] - \citet{jensen03},
[14] - \citet{nurnberger98},
[15] - \citet{beckwith90},
[16] - \citet{duvert00},
[17] - \citet{kitamura02},
[18] - \citet{pietu06},
[19] - \citet{lommen10},
[20] - \citet{weintraub89},
[21] - \citet{beckwith91}, 
[22] - \citet{dutrey98},
[23] - \citet{looney00},
[24] - \citet{hughes09},
[25] - \citet{isella09},
[26] - \citet{guilloteau98}.}
\end{deluxetable}

\end{document}